\documentclass[final,times,twocolumn]{elsarticle}
\usepackage{graphicx,amssymb,amsmath,hyperref}
\usepackage{multirow,dcolumn,bm,latexsym,soul}

\journal{Physics of the Dark Universe}
\begin{document}
\newcommand{\be}{\begin{equation}}
\newcommand{\ee}{\end{equation}}
\newcommand{\bq}{\begin{eqnarray}}
\newcommand{\eq}{\end{eqnarray}}
\begin{frontmatter}

\title{Current and future cosmological impact of microwave background temperature measurements}
\author[inst1,inst2]{C. J. A. P. Martins\corref{cor1}}\ead{Carlos.Martins@astro.up.pt}
\author[inst1]{A. M. M. Vieira}\ead{ana.mafalda.vieira@sapo.pt}
\address[inst1]{Centro de Astrof\'{\i}sica da Universidade do Porto, Rua das Estrelas, 4150-762 Porto, Portugal}
\address[inst2]{Instituto de Astrof\'{\i}sica e Ci\^encias do Espa\c co, CAUP, Rua das Estrelas, 4150-762 Porto, Portugal}
\address[inst3]{Faculdade de Ci\^encias da Universidade de Lisboa, Campo Grande, 1749-016 Lisboa, Portugal}
\cortext[cor1]{Corresponding author}

\begin{abstract}
{The redshift dependence of the cosmic microwave background temperature, $T(z)=T_0(1+z)$,  is a key prediction of standard cosmology, but this relation is violated in many extensions thereof. Current astrophysical facilities can probe it in the redshift range $0\le z\le6.34$. We extend recent work by Gelo {\it et al.} (2022) showing that for several classes of models (all of which aim to provide alternative mechanisms for the recent acceleration of the universe) the constraining power of these measurements is comparable to that of other background cosmology probes. Specifically, we discuss constraints on two classes of models not considered in the earlier work: a model with torsion and a recently proposed phenomenological dynamical dark energy model which can be thought of as a varying speed of light model. Moreover, for both these models and also for those in the earlier work, we discuss how current constraints may be improved by next-generation ground and space astrophysical facilities. Overall, we conclude that these measurements have a significant cosmological impact, mainly because they often constrain combinations of model parameters that are orthogonal, in the relevant parameter space, to those of other probes.}
\end{abstract}
\end{frontmatter}

\section{Introduction}
\label{introd}

One of the best-known properties of our universe is the present-day value of the cosmic microwave background (CMB) temperature, which is $T_0= 2.7255\pm 0.0006$ K \cite{Fixsen}. Considerably less known is the fact that, assuming that the CMB spectrum was originally a black-body, the expansion of the Universe is adiabatic, and the photon number is conserved, the CMB temperature will evolve as
\be
T(z)=T_0(1+z)\,.
\ee
While this holds true in the standard cosmological model, it does not in many extensions thereof, and several possible physical mechanisms can be responsible for deviations from it \cite{Avgoustidis,Euclid}. Observational evidence for deviations from this relation will  constitute evidence for new physics, which would need to be subsequently characterized by additional observational and experimental probes. This provides a motivation for measurements of this temperature at the broadest possible range of redshifts.

At present, such measurements can be obtained using two different observational techniques. At lower redshifts (typically $z<1$), one relies on the the thermal Sunyaev-Zel'dovich (SZ) effect, while in the approximate range $1<z<3$ one can rely on high-resolution spectroscopy of suitable molecular or atomic species whose energy levels can be excited by CMB photons. After several decades of obtaining upper limits, the first such measurement was obtained by Srianand {\it et al.} \cite{Srianand}, and more that two further decades of effort on this subject have led to a set of several tens of measurements, with a redshift range which has recently been enlarged to reach $z\sim6.34$ \cite{Riechers}.

While these measurements have occasionally been used to constrain specific models, a recent work \cite{Gelo} has provided a first comparative assessment of the cosmological constraining power of this data, specifically comparing it to that of other commonly used low redshift background data sets. That work focused on three different model classes: the decaying dark energy model of Jetzer {\it et al.} \cite{Jetzer1}, the scale invariance model of Canuto {\it et al.} \cite{Canuto1}, and the recently proposed fractional cosmology model \cite{Fractional}. The main result emerging from that analysis is that, broadly speaking, the constraining power of the CMB temperature measurements is comparable to that of other low redshift cosmological data sets, with the caveat that there are model dependencies, ranging from cases where these measurements provide a $10\%$ to $20\%$ improvement in constraints to those where they rule out otherwise viable models.

For our present purposes, all models beyond the canonical $\Lambda$CDM can be divided into two main classes. The first class comprises models which are genuine parametric extensions of $\Lambda$CDM, in the sense the model reduces to $\Lambda$CDM for some particular choice of its parameters. Sometimes one of these parameters is (or can be rephrased as) a dark energy equation of state, and this parameter tends to also impact the temperature-redshift relation. The second class comprises models which, nominally at least, have no parametric $\Lambda$CDM limit, relying instead on a different physical mechanism to account for the recent acceleration of the universe. However, they can also be turned into parametric extensions of $\Lambda$CDM by allowing for the possibility of a non-zero cosmological constant. If so these models will contain two possible mechanisms for acceleration, which may still be observationally distinguishable, although they may well be strongly degenerate, For models in this second class the case without a cosmological constant is typically ruled out by observations, so only the latter case is worth considering. Examples of this can be seen in \cite{Gelo,Marques}, and we will encounter additional ones in what follows.

Here we extend the work of Gelo {\it et al.} \cite{Gelo} in two different ways. Firstly we carry out analogous studies for two other model classes, which stem from physical motivations which differ from those of the three previously studied ones: a model including torsion \cite{Torsion1,Torsion2} and a recent phenomenological dynamical dark energy model \cite{Gupta}, which can be thought of as a varying speed of light model and turns out to have various similarities with some of the other models under consideration. Secondly, for all five models, we go beyond currently available data and discuss how current constraints may be improved by hypothetical next-generation ground and space astrophysical facilities providing, among others, improved measurements of the CMB temperature. Our analysis leads us to expect that these measurements will continue to have a significant cosmological impact in the next decade. Throughout this work we use units with $c=\hbar=1$.

\section{Current and future data}
\label{datacf}

In this section we briefly describe the current (publicly available) background cosmology data and also the analogous future (simulated) data which is used in what follows. In the subsequent sections this data will be used in standard likelihood analyses, as described e.g. in \cite{Verde}, for several different cosmological models. 

The likelihood is defined as
\be
{\cal L}(p)\propto\exp{\left(-\frac{1}{2}\chi^2(p)\right)}\,,
\ee
where $p$ symbolically denotes the free parameters in the model being considered. The chi-square for a relevant redshift-dependent quantity $E(z)$ has the explicit form
\be
\chi^2(p)=\sum_{i,j}\left(E_{obs,i}-E_{mod,i}(p)\right)C_{ij}^{-1}\left(E_{obs,j}-E_{mod,j}(p)\right)\,,
\ee
where the obs and mod subscripts denote observations and model respectively, and $C$ is the covariance matrix of the dataset (which may, in some cases, be trivial).

Our analyses are grid-based, and unless otherwise is stated we use uniform priors for the various model parameters, in the plotted ranges. We have tested that these assumptions do not impact our results, Two independent codes have been used (one written in Matlab, the other in Python), validated and verified against each other and previous results in the literature.

\subsection{Current data}
\label{nowdata}

The existing CMB temperature measurements are listed in Tables \ref{table1} and \ref{table2} for the reader's convenience. They were also used for cosmological purposes in the aforementioned work \cite{Gelo}. The first of these tables contains the low redshift measurements, coming from the thermal SZ effect. Specifically these come from a set of 815 Planck clusters organized into 18 redshift bins as discussed in \cite{Hurier}, as well as a sample of 158 SPT clusters organized in 12 redshift bins as discussed in \cite{Saro}. We refer the reader to these works for further details on how each of the samples has been acquired and on how the data analysis leading to the quoted measurements was done.

The second table contains the (usually higher redshift) spectroscopic measurements, which are obtained from observations at various wavelengths (from optical to radio) and using several molecular or atomic species. Part of the data comes from a recent compilation \cite{Klimenko}, which has updated several earlier analyses \cite{Srianand0,Noterdaeme1,Noterdaeme2,Krogager}, specifically by accounting for the contribution of collisional excitation in the diffuse interstellar medium to the excitation temperature of the tracer species. The highest redshift measurement currently available (though one providing a relatively weak constraint) is at $z\sim6.34$, but most of them are at considerably lower redshifts, $z\sim2.5$. In passing, we note that spectroscopic methods also provide some additional upper limits on $T(z)$, at comparable redshifts; a recent compilation can be found in \cite{Klimenko}. We do not include these upper limits since they would carry negligible statistical weight in our analysis.

\begin{table}
\begin{center}
\caption{CMB temperature measurements, with 1-$\sigma$ statistical uncertainties, obtained from SZ measurements.}
\label{table1}
\begin{tabular}{c c c}
\hline
Redshift & $T_{CMB}(z)$ [K] & Reference \\
\hline
$0.037$ & $2.888\pm0.041$ & \cite{Hurier} \\
$0.072$ & $2.931\pm0.020$ & \cite{Hurier} \\
$0.125$ & $3.059\pm0.034$ & \cite{Hurier} \\
$0.129$ & $3.01_{-0.11}^{+0.14}$ & \cite{Saro} \\
$0.171$ & $3.197\pm0.032$ & \cite{Hurier} \\
$0.220$ & $3.288\pm0.035$ & \cite{Hurier} \\
$0.265$ & $3.44_{-0.13}^{+0.16}$ & \cite{Saro} \\
$0.273$ & $3.416\pm0.040$ & \cite{Hurier} \\
$0.332$ & $3.562\pm0.052$ & \cite{Hurier} \\
$0.371$ & $3.53_{-0.14}^{+0.18}$ & \cite{Saro} \\
$0.377$ & $3.717\pm0.065$ & \cite{Hurier} \\
$0.416$ & $3.82_{-0.15}^{+0.19}$ & \cite{Saro} \\
$0.428$ & $3.971\pm0.073$ & \cite{Hurier} \\
$0.447$ & $4.09_{-0.19}^{+0.25}$ & \cite{Saro} \\
$0.471$ & $3.943\pm0.113$ & \cite{Hurier} \\
$0.499$ & $4.16_{-0.20}^{+0.27}$ & \cite{Saro} \\
$0.525$ & $4.380\pm0.120$ & \cite{Hurier} \\
$0.565$ & $4.075\pm0.157$ & \cite{Hurier} \\
$0.590$ & $4.62_{-0.26}^{+0.36}$ & \cite{Saro} \\
$0.619$ & $4.404\pm0.195$ & \cite{Hurier} \\
$0.628$ & $4.45_{-0.23}^{+0.31}$ & \cite{Saro} \\
$0.676$ & $4.779\pm0.279$ & \cite{Hurier} \\
$0.681$ & $4.72_{-0.27}^{+0.39}$ & \cite{Saro} \\
$0.718$ & $4.933\pm0.371$ & \cite{Hurier} \\
$0.742$ & $5.01_{-0.33}^{+0.49}$ & \cite{Saro} \\
$0.783$ & $4.515\pm0.621$ & \cite{Hurier} \\
$0.870$ & $5.356\pm0.617$ & \cite{Hurier} \\
$0.887$ & $4.97_{-0.19}^{+0.24}$ & \cite{Saro} \\
$0.972$ & $5.813\pm1.025$ & \cite{Hurier} \\
$1.022$ & $5.37_{-0.18}^{+0.22}$ & \cite{Saro} \\
\hline
\end{tabular}
\end{center}
\end{table}
\begin{table}
\begin{center}
\caption{CMB temperature measurements, with 1-$\sigma$ statistical uncertainties, obtained from spectroscopic measurements.}
\label{table2}
\begin{tabular}{c c c}
\hline
Redshift & $T_{CMB}(z)$ [K]  & Reference \\
\hline
$0.89$ & $5.08\pm0.10$ & \cite{Muller} \\
$1.73$ & $7.9^{+1.7}_{-1.4}$ & \cite{Klimenko} \\ 
$1.77$ & $6.6^{+1.2}_{-1.1}$ & \cite{Klimenko} \\ 
$1.78$ & $7.2\pm0.8$ & \cite{Cui} \\
$1.97$ & $7.9\pm1.0$ & \cite{Ge} \\ 
$2.04$ & $8.6^{+1.9}_{-1.4}$ & \cite{Klimenko}  \\ 
$2.34$ & $10\pm4$ & \cite{Srianand} \\
$2.42$ & $9.0^{+0.9}_{-0.7}$ & \cite{Klimenko} \\ 
$2.53$ & $9.8^{+0.7}_{-0.6}$ & \cite{Klimenko} \\ 
$2.63$ & $10.8^{+1.4}_{-3.3}$ & \cite{Klimenko} \\ 
$2.69$ & $10.4^{+0.8}_{-0.7}$ & \cite{Klimenko} \\ 
$3.02$ & $12.1^{+1.7}_{-3.2}$ & \cite{Molaro} \\ 
$3.09$ & $12.9^{+3.3}_{-4.5}$ &\cite{Klimenko} \\ 
$3.29$ & $15.2^{+1.0}_{-4.2}$ & \cite{Klimenko} \\  
$6.34$ & $23.1^{+7.1}_{-6.7}$ & \cite{Riechers} \\
\hline
\end{tabular}
\end{center}
\end{table}

Since, in our context, the CMB temperature measurements are effectively background cosmology data, our assessment of their constraining power will be done against two other background cosmology data sets. The first is the Pantheon compilation \cite{Scolnic,Riess}. This is originally a 1048 supernova data set, subsequently compressed\footnote{Strictly speaking, this compression also relies on 15 Type Ia supernovae from two Multi-Cycle Treasury programs, and it assumes a spatially flat universe. The compression methodology and validation are detailed in Section 3 of  \cite{Riess}.} into 6 correlated measurements of $E^{-1}(z)$ (where $E(z)=H(z)/H_0$ is the dimensionless Hubble parameter) in the redshift range $0.07<z<1.5$. The second is the compilation of 38 Hubble parameter measurements of Farooq {\it et al.} \cite{Farooq}: this is a more heterogeneous set includes both data from cosmic chronometers and from baryon acoustic oscillations. Both of these are canonical data sets and have been extensively used in the literature, including \cite{Gelo}. We retain them in the present work, to enable a fair comparison to the earlier one. Together, the two cosmological data sets contain measurements up to redshift $z\sim2.36$ (thus a redshift range comparable to that of the CMB measurements), and when using the two in combination we will refer tho this as the cosmological data.

Note that in all our analyses the Hubble constant is never used as a free parameter; instead it is always analytically marginalized, following the procedure described in \cite{Anagnostopoulos}, so our results do not depend on possible choices of this parameter. Since one can trivially write $H(z)=H_0 E(z)$, one notices that $H_0$ is purely a multiplicative constant, and can be analytically integrated in the likelihood. Towards this end one computes quantities
\bq
A(p)&=&\sum_{i}\frac{E_{mod,i}^2(p)}{\sigma^2_i}\\
B(p)&=&\sum_{i}\frac{E_{mod,i}(p) H_{obs,i}}{\sigma^2_i}\\
C(p)&=&\sum_{i}\frac{H_{obs,i}^2}{\sigma^2_i}
\eq
where the $\sigma_i$ are the uncertainties in observed values of the Hubble parameter. Then, the chi-square is given by
\be
\chi^2(q)=C(q)-\frac{B^2(q)}{A(q)}+\ln{A(q)}-2\ln{\left[1+Erf{\left(\frac{B(q)}{\sqrt{2A(q)}}\right)}\right]}
\ee
where $Erf$ is the Gauss error function and $\ln$ is the natural logarithm. The same remark holds for the present-day value of the CMB temperature, $T_0$, which is also analytically marginalized,

In passing, we note that although measurements of the CMB temperature cannot {\it per se}, address the Hubble tension, they may indirectly help shed some light on it by improving constraints on cosmological model parameters degenerate with $H_0$.

\subsection{Simulated future data}
\label{newdata}

In this work we also present forecasts for future constraints on the various models under consideration. For this we will replace the data discussed in the previous subsection by three analogous simulated ones, which represent (in an admittedly simplified way) the expected performance of several next-generation astrophysical facilities---some under construction, others merely proposed. We will proceed by generating mock data, always including Gaussian noise, and using the same analysis pipelines providing the current constraints to obtain the forecast constraints.

For generating all this mock data, a flat $\Lambda$CDM fiducial model, with a matter density $\Omega_m=0.3$ and a Hubble constant $H_0=70$ km/s/Mpc, is assumed throughout. This choice stems from the fact that there is currently no substantive evidence of deviations from this. One could simulate additional non-$\Lambda$CDM cases for each model, but that would make the work much longer, without necessarily providing much additional insight. While the degeneracies between model parameters imply that the expected future parameter constraints will depend on the fiducial values, such dependencies are expected to be small since any plausible models must be (in a loose sense) close to $\Lambda$CDM in parameter space, at least for low redshifts.

Starting with the CMB temperature and the SZ measurements, the data in Table \ref{table1} is replaced by an assumed sample of high signal-to-noise clusters from a next-generation space mission such as CORE \cite{CORE}. For simplicity we assume them to be uniformly spaced in 20 bins in the redshift range $z\in[0.05,1.00]$, with an absolute uncertainty in the temperature ranging from 0.005 K to 0.05 K at the lower and higher redshifts respectively.

As for the spectroscopic measurements in Table \ref{table2}, considering the difficulty of finding additional lines of sight where these measurements can be made, we conservatively assume a data set consisting of the 13 absorption systems for which the measurements are done in the optical/UV (that is, all except the ones at lowest and highest redshift), but with an improved absolute temperature uncertainty of 0.1 K, as one may expect with the ANDES high-resolution spectrograph at the ELT, formerly known as ELT-HIRES \cite{HIRES,ANDES}. For context, we note that one can see in Table \ref{table2} that current best constraints of this type have uncertainties of about 0.7 K, and that measurements with ESPRESSO, which recently became operational at the VLT, is expected to reach sensitivities of ca. 0.4 to 0.5 K.

Regarding cosmological data, as a replacement for the Pantheon compilation we assume a future data set of supernova measurements from the proposed Roman Space Telescope, formerly known as WFIRST. Their preliminary study of these authors \cite{Riess,WFIRST} reports the following values for percent errors on the dimensionless Hubble parameter $E(z)$:  $\sigma_e=1.3, 1.1, 1.5, 1.5, 2.0, 2.3, 2.6, 3.4, 8.9$ for the nine redshift bins centered at redshifts $z=0.07, 0.20, 0.35, 0.60, 0.80, 1.00, 1.30, 1.70, 2.50$, respectively. While these measurements are not fully independent, the authors also state that pairwise correlations between the measurements are small (and are not explicitly provided); for simplicity we therefore ignore them in the analysis that follows. We further note that their simulated data set was also obtained under the assumption of a flat universe, but this is compatible with our choice of fiducial model.

Finally, as replacement for the Hubble parameter measurements compilation of \cite{Farooq}, we will take a future set of measurements of the redshift drift \cite{Sandage,Alves,Rocha}, expected to be carried out by ANDES at high redshifts \cite{Liske,HIRES,ANDES} and by the SKA at lower redshift \cite{Klockner}, This is a direct model-independent probe  of the expansion of the universe; the redshift drift of an astrophysical object following the Hubble flow is given by \cite{Sandage}
\be
\Delta z=\tau_{exp} H_0 \left[1+z-E(z)\right]\,,
\ee
where $\tau_{exp}$ is the experiment time span (not to be confused with the on-sky observation time), although the actual observable is a spectroscopic velocity
\be
\Delta v=k\tau_{exp}h\left[1-\frac{E(z)}{1+z}\right]\,,
\ee
which, for subsequent convenience, we expressed in terms of $E(z)$ and $h=H_0/(100\,  km/s/Mpc)$; we also introduced the normalization constant $k$, which has the value $k=3.064$ cm/s if $\tau_{exp}$ is expressed in years. For the SKA, following the results in \cite{Klockner}, we assume a set of 10 measurements, equally spaced between redshifts $z = 0.1$ and $z = 1.0$ and with the associated spectroscopic velocity uncertainties equally spaced between $1\%$ and $10\%$ respectively. For ANDES we assume the recently proposed Golden Sample of \cite{Cristiani}, consisting of seven measurements with spectroscopic velocity uncertainties $\sigma_v=10.4, 12.5, 11.1, 9.8, 7.8, 11.2, 12.1$ at redshifts $z=2.999, 3.240, 3.535, 3.897, 3.960, 4.147, 4.760$, respectively.

\section{Steady-state torsion models}
\label{torsion}

From a mathematical (geometrical) perspective, one of the most natural extensions of Einstein's General Theory of Relativity is the inclusion of space-time torsion \cite{Cartan,Kibble,Sciama}. It is moreover possible to choose the torsion tensor (defined as the antisymmetric part of the affine connection) such that the homogeneity and isotropy of Friedmann-Lemaitre-Robertson-Walker (FLRW) universes is preserved, as shown by \cite{Tsamparlis}. Following suggestions that such models might explain the recent acceleration of the universe \cite{Torsion1,Torsion2}, under the so-called steady-state torsion assumption of a constant fractional contribution of torsion to the volume expansion, constraints on these models have been obtained from the Type Ia Supernova and Hubble parameter measurements introduced in Sect. \ref{nowdata} \cite{Marques}. In summary, these show that such models are an example of the second class of models described in the introduction. Although they were originally proposed as genuine alternatives to $\Lambda$CDM  (in the sense that torsion alone would be expected to yield the current acceleration of the universe), that scenario is observationally ruled out. Instead if one treats the models as one-parameter extensions of $\Lambda$CDM, by allowing for the presence of a cosmological constant, the relative contribution of torsion is constrained to the level of a few percent of the energy density budget. Here we show how these constraints are improved by the addition of CMB temperature measurements.

The general field equations including torsion are known as the Einstein-Cartan equations. In order to obtain homogeneous and isotropic universes the torsion tensor must depend on a scalar function $\phi$ which depends only on time but is otherwise arbitrary. Under these assumptions one obtains the following Friedmann, Raychaudhuri and continuity equations \cite{Torsion1,Torsion2}
\be
H^2=\frac{1}{3}\kappa\rho-\frac{K}{a^2}+\frac{1}{3}\Lambda-4\phi^2-4H\phi
\ee
\be
\frac{\ddot a}{a}=-\frac{1}{6}(\rho+3p)+\frac{1}{3}\Lambda-2{\dot\phi}-2H\phi
\ee
\be
{\dot\rho}=-3H\left(1+2\frac{\phi}{H}\right)(\rho+p)+4\phi\left(\rho+\frac{\Lambda}{\kappa}\right)\,,
\ee
where $\kappa=8\pi G$; in what follows we will assume $K=0$ and a barotropic fluid with a constant equation of state $p=w\rho$; we note that this is the equation of state of matter, so the canonical case corresponds to $w=0$.

It is convenient to introduce the canonical present-day fractions of matter and dark energy, $\Omega_m=\kappa\rho_{0}/3H_0^2$ and $\Omega_\Lambda=\Lambda/3H_0^2$, and we can analogously define a torsion contribution
\be\label{torphi}
\Omega_\phi=-4\left(\frac{\phi_0}{H_0}\right)\left[1+\frac{\phi_0}{H_0}\right]\,.
\ee
In steady-state torsion models \cite{Torsion1} one further assumes that the relative torsion contribution to the expansion remains constant in time,
\be
\frac{\phi}{H}=\lambda=const.\,,
\ee
and one can therefore express observational constraints in terms of the dimensionless model parameter $\lambda$, the two parameters being related via $\Omega_\phi=-4\lambda(1+\lambda)$.

For numerical purposes, it is also convenient to define a dimensionless density, denoted $r$, via $\rho=r\rho_0$, and to write the Friedmann and continuity equations as a function of redshift as follows
\be
E^2(z)=1+\frac{r(z)-1}{(1+2\lambda)^2}\Omega_m
\ee
\be
(1+z)\frac{dr}{dz}=3(1+w)r+2\lambda\left[2+(1+3w)r-\frac{2}{\Omega_m}(1+2\lambda)^2\right]\,,
\ee
where we have made use of the flatness condition (leading to a relation between $\lambda$, $\Omega_m$ and $\Omega_\Lambda$ which allows us to eliminate the latter) and are also making the quite safe assumptions that $\lambda\neq-1/2$ and $\Omega_m\neq0$. Last but not least, we have also neglected the radiation density, since we are dealing with low-redshift data.

Importantly, the CMB temperature data also constrains this model. As shown by \cite{Bolejko},  in this model one has a violation of the distance duality relation (which relates the luminosity and angular diameter distances), given by
\be
\eta\equiv\frac{D_l}{(1+z)^2D_A}-1=\lambda \ln{(1+z)}\,.
\ee
Following \cite{Avgoustidis11}, and assuming adiabatic processes, we therefore expect the corresponding violation of the temperature-redshift relation to have the form
\be
T(z)=T_0(1+z)\left[1+\lambda \ln{(1+z)}\right]^{2/3}\,;
\ee
note that this is independent of the matter density and depends only on $\lambda$. It can therefore be used to break the degeneracy between this parameter and the matter density.

\begin{table*}
\begin{center}
\caption{Current constraints and forecasts for the steady-state torsion model, under various assumptions on the model parameters and data. Cosmo refers to the current Type Ia supernova and Hubble parameter data, CMB to the current temperature data, and Forecast to a combination of future data sets. The top four rows are the results of \cite{Marques}, and are reproduced here to facilitate the comparison to our updated constraints, which are shown in the next four rows; the last two rows show our forecasts.}
\label{table3}
\begin{tabular}{c c c c c c}
\hline
Data & Assumptions & $\Omega_m$ & $\lambda$ & $w$ & Reference \\
\hline
Cosmo & $w=0$, No $\Omega_m$ prior & $0.18^{+0.06}_{-0.03}$ & $-0.07^{+0.05}_{-0.04}$ & N/A & \cite{Marques}\\
Cosmo & $w=0$, Planck $\Omega_m$ prior & Recovers prior & $0.02^{+0.01}_{-0.02}$ & N/A  & \cite{Marques}\\ 
Cosmo & $w=const.$, No $\Omega_m$ prior & Unconstrained & $-0.02\pm0.04$ & $-0.05\pm0.04$ & \cite{Marques}\\
Cosmo & $w=const.$, Planck $\Omega_m$ prior & Recovers prior & $-0.01\pm0.02$ & $-0.05\pm0.03$ & \cite{Marques}\\ 
\hline
Cosmo + CMB & $w=0$, No $\Omega_m$ prior & $0.24\pm0.04$ & $-0.03\pm0.02$ & N/A & This work\\
Cosmo + CMB & $w=0$, Planck $\Omega_m$ prior & Recovers prior & $0.01\pm0.01$ & N/A & This work\\ 
Cosmo + CMB & $w=const.$, No $\Omega_m$ prior & $0.30^{+0.08}_{-0.07}$ & $-0.015\pm0.025$ & $-0.05^{+0.04}_{-0.03}$ & This work\\
Cosmo + CMB & $w=const.$, Planck $\Omega_m$ prior & Recovers prior & $-0.010\pm0.015$ & $-0.05^{+0.03}_{-0.02}$ & This work\\ 
\hline
Forecast & $w=const.$, No $\Omega_m$ prior & $0.31\pm0.01$ & $-0.001\pm0.004$ & $-0.02\pm0.02$ & This work\\
Forecast & $w=const.$, Planck $\Omega_m$ prior & Recovers prior & $-0.001\pm0.003$ & $-0.01\pm0.01$ & This work\\ 
\hline
\end{tabular}
\end{center}
\end{table*}

\begin{figure*}
\begin{center}
\includegraphics[width=0.32\textwidth]{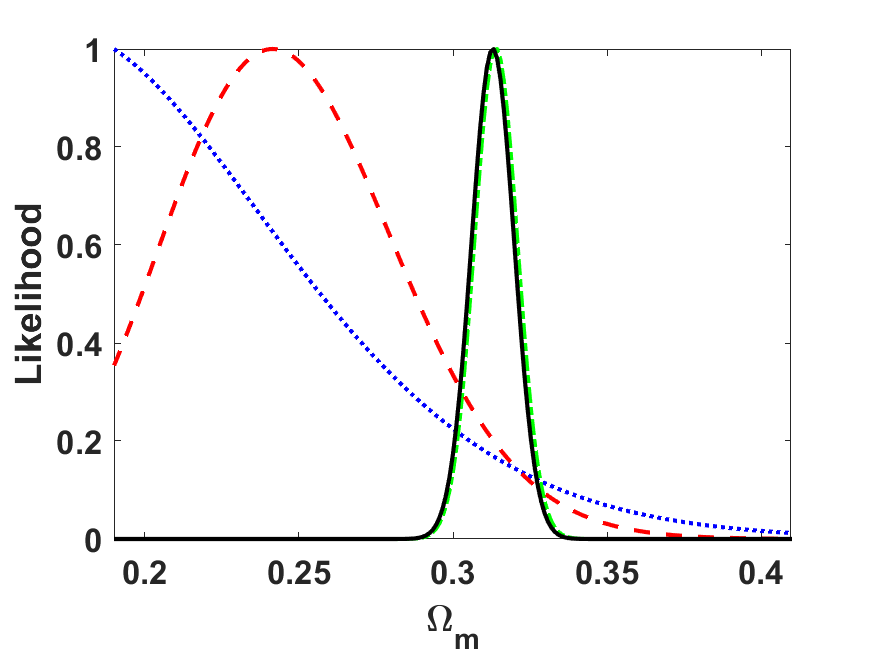}
\includegraphics[width=0.32\textwidth]{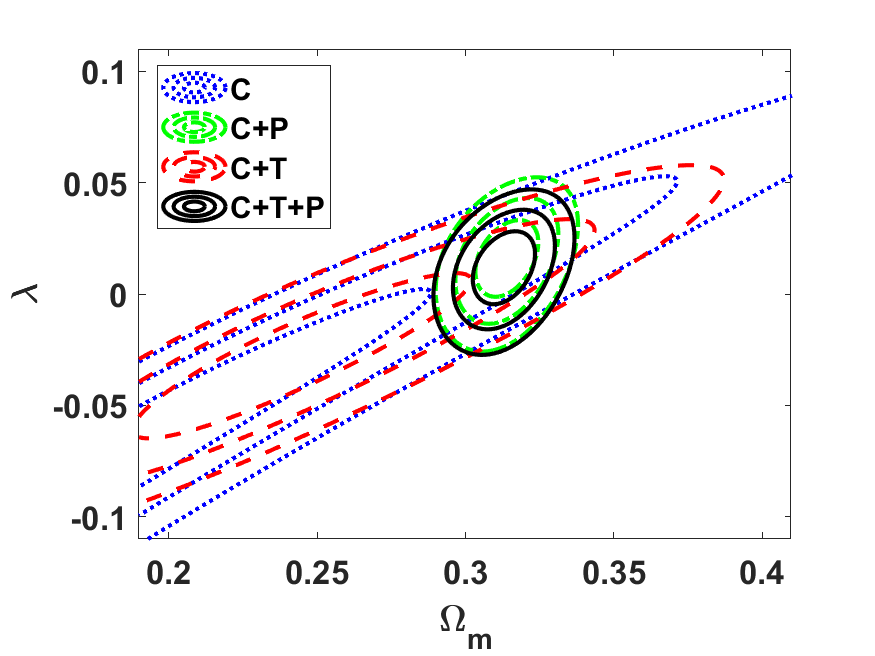}
\includegraphics[width=0.32\textwidth]{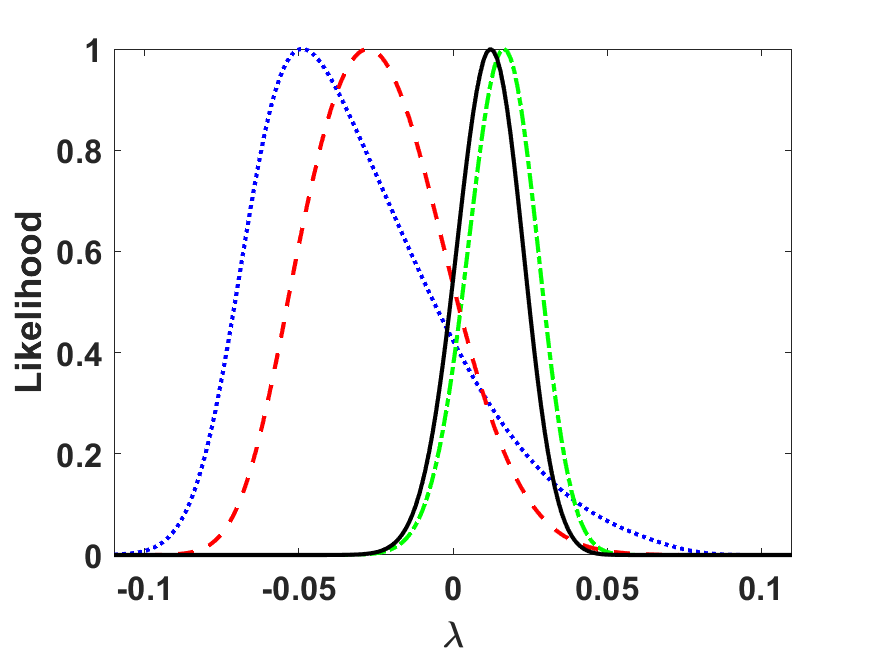}
\end{center}
\caption{Current constraints on steady-state torsion with standard matter, $w=0$. The middle panel shows the one, two and three sigma constraints in the two-dimensional $\Omega_m$--$\lambda$ parameter space, and the side panels show the one-dimensional (marginalized) posterior likelihoods for each of the parameters. Blue dotted, green dash-dotted, red dashed and black solid curves depict the results for the Cosmology, Cosmology + Planck prior, Cosmology + CMB, and Cosmology + CMB + Planck prior data respectively.}
\label{figure01}
\end{figure*}
\begin{figure*}
\begin{center}
\includegraphics[width=0.32\textwidth]{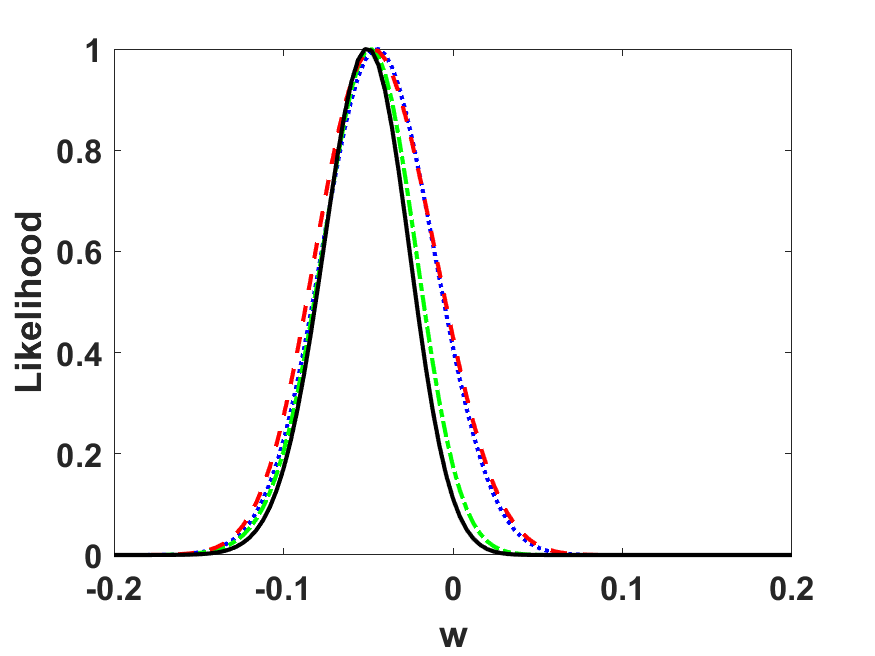}
\includegraphics[width=0.32\textwidth]{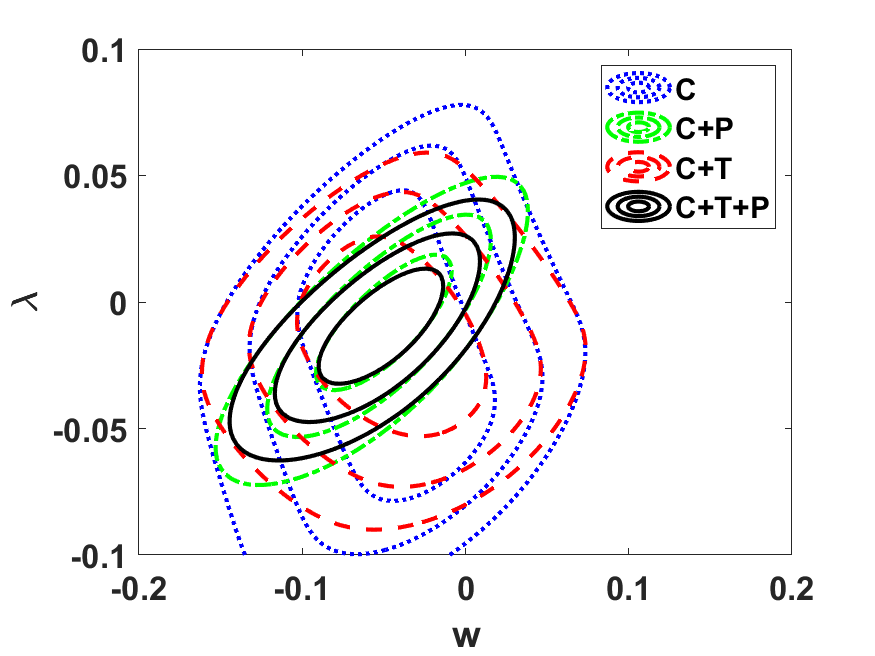}
\includegraphics[width=0.32\textwidth]{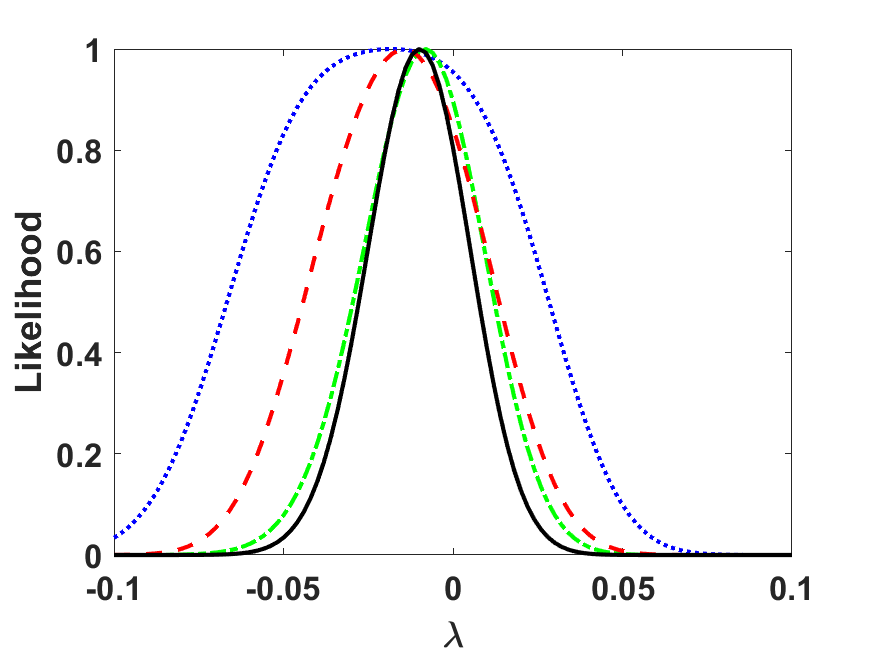}
\end{center}
\caption{Current constraints on steady-state torsion with $w=const$ (but not necessarily vanishing) and $\Omega_m$ marginalized. The middle panel shows the one, two and three sigma constraints in the two-dimensional $w$--$\lambda$ parameter space, and the side panels show the one-dimensional (marginalized) posterior likelihoods for each of the parameters. Blue dotted, green dash-dotted, red dashed and black solid curves depict the results for the Cosmology, Cosmology + Planck prior, Cosmology + CMB, and Cosmology + CMB + Planck prior data respectively.}
\label{figure02}
\end{figure*}

\subsection{Current constraints}

The top part of Table \ref{table3} summarizes the constraints on the model reported by \cite{Marques} under various assumptions on the model parameters. These were obtained using the supernova and Hubble parameter data discussed in Sect. \ref{nowdata} (which we collectively refer to as cosmological data), sometimes complemented by a Planck-like prior on the matter density, $\Omega_m=0.315\pm0.007$ \cite{Planck}; for all the other parameters, uniform priors were used.

We have repeated this analysis, with and without the addition of the CMB temperature, the latter being a simple means of validating the code by reproducing the results of the earlier work (although the two codes are not fully independent). The new results are summarized in the middle part of table \ref{table3}, and show a discernible effect of the addition of the CMB data.

Figure \ref{figure01} shows relevant results for the conservative case where matter is assumed to have the standard equation of state, $w=0$. Without the Planck prior, the most salient aspect is that there is a degeneracy between the matter density and the dimensionless torsion parameter $\lambda$, which allows for a reduced matter density (by comparison to its standard value), compensated by a slightly negative value of $\lambda$. The addition of the CMB temperature data strongly restricts this possibility, improving the constraint on $\lambda$ by more than a factor of two. Naturally, the Planck prior on $\Omega_m$ is an alternative way to break this degeneracy (and in that case the constraint on the matter density simply recovers the prior), but even in this case the CMB temperature data provides a slight improvement in the $\lambda$ constraint.

Figure \ref{figure02} shows the analogous results for the more general case where matter is allowed to have a non-zero (but constant) equation of state. Qualitatively the results are the same, though one should note that in this case the extended parameter space implies that, in the absence of the Planck prior and the CMB temperature data, the matter density is fully unconstrained. The addition of the CMB data changes this, and enables a non-trivial (though still comparatively weak) constraint to be obtained. The constraints on $\lambda$ and $w$ are also improved by the CMB data. In the former case this is a direct effect, since the CMB temperature is explicitly dependent on $\lambda$; in the latter it is an indirect and therefore weaker effect, due to the partial breaking of degeneracies with the other parameters. We also note that the constraints on $w$ are qualitatively comparable to those in previous analogous works, such as \cite{Tutusaus}, although no direct quantitative comparison is possible since the two works address different types of models.

\begin{figure*}
\begin{center}
\includegraphics[width=0.32\textwidth]{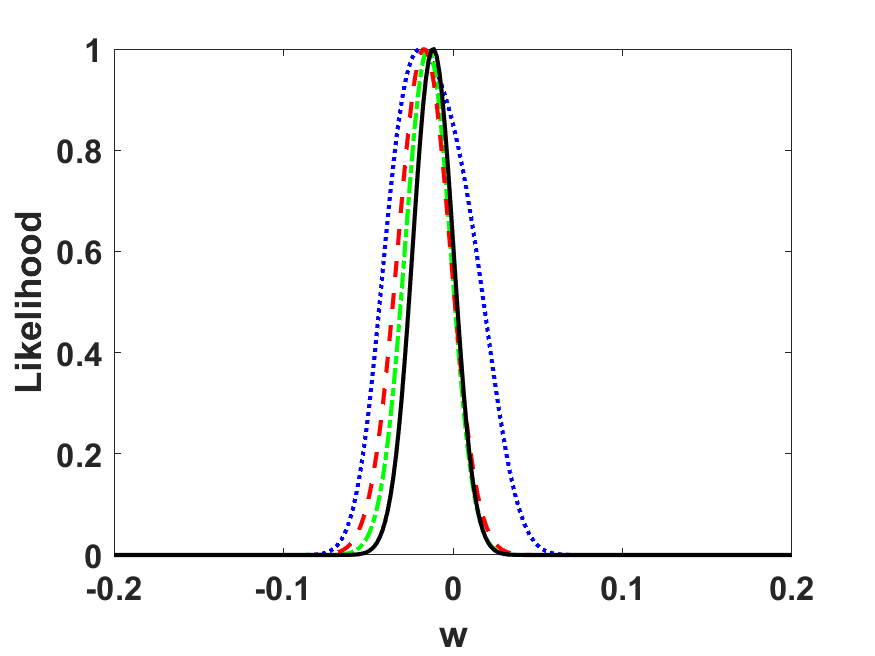}
\includegraphics[width=0.32\textwidth]{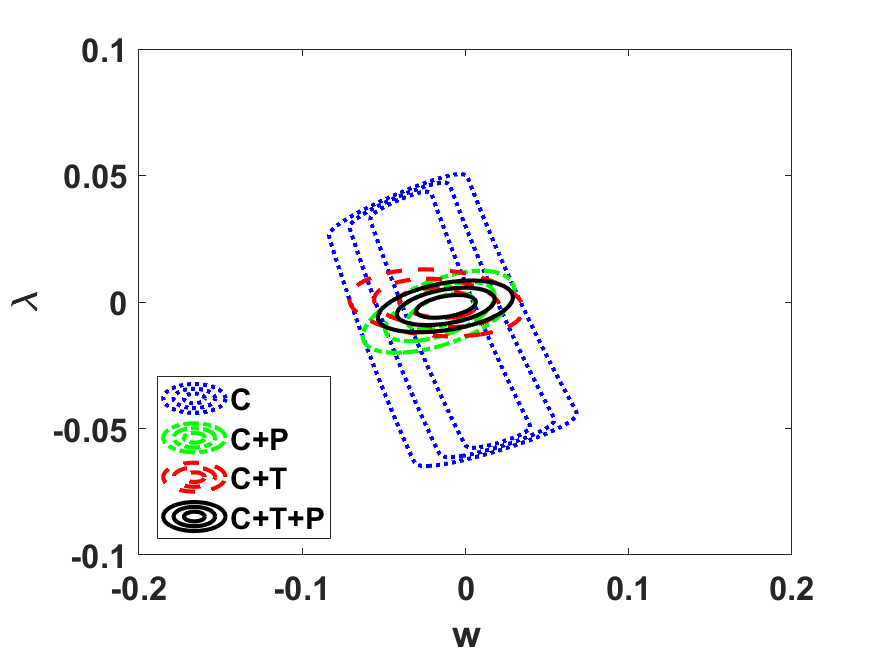}
\includegraphics[width=0.32\textwidth]{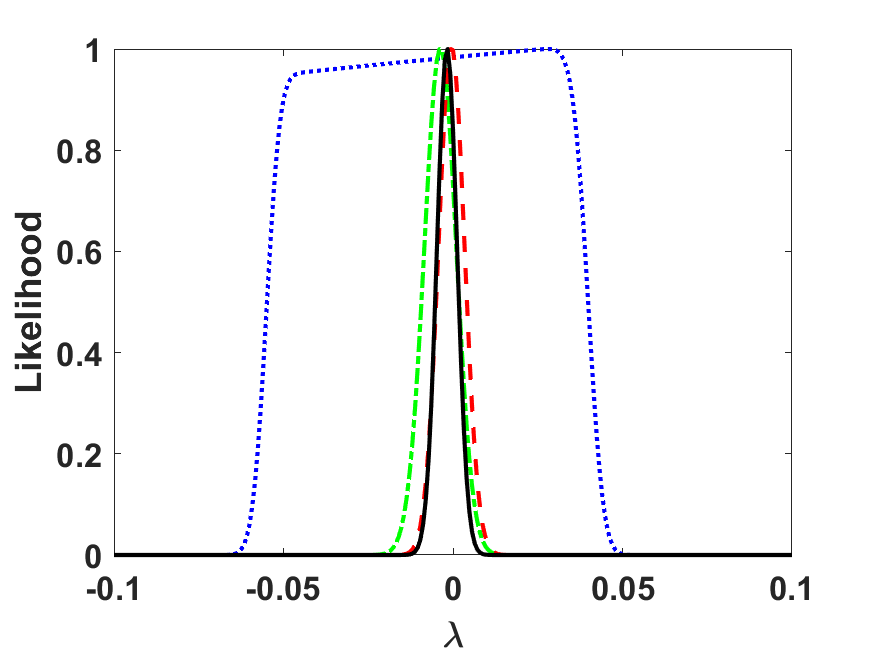}
\end{center}
\caption{Same as Figure \ref{figure02}, for the simulated future data discussed in the text.}
\label{figure03}
\end{figure*}

\subsection{Forecasts}

We now repeat the analysis of the full three-dimensional parameter space of the model, ($\Omega_m,\lambda,w$), using instead the future simulated data described in Sect. \ref{newdata}. Here we also include a Planck-like prior with and uncertainty $\sigma_{\Omega_m}=0.007$. We start by noting that \cite{Marques} already provided some forecasts, though under slightly different assumptions: their future supernova data is the same as ours, but their assumptions on the redshift drift are different (they used no SKA data, and different ELT-ANDES data), and they used no CMB temperature data. With a Planck-level prior on the matter density, their forecast provided the following one-sigma uncertainties
\bq
\sigma_\lambda&=&0.009\\
\sigma_w&=&0.013\,;
\eq
the former corresponds to a one-sigma upper bound on the torsion contribution of $\Omega_\phi<0.036$. In the present work, in addition to using CMB temperature and SKA redshift drift data, we also update the ELT redshift drift data, relying on the recently proposed Golden Sample of \cite{Cristiani}.

Figure \ref{figure03} and the bottom part of Table \ref{table3} summarize the results of our new forecast. This confirms the constraining power of the CMB temperature data, which improves the constraints on $\lambda$ by about a factor of 3 with respect to the cosmological data (without these measurements) and by a factor of 6 with respect to current constraints. One may notice that the posterior distribution of $\lambda$ for the cosmology data differs from the analogous one in Fig. \ref{figure02}. The reason for this is that the current $H(z)$ dataset is replaced in our forecasts by a set of redshift drift measurements, which are analogous to (but not exactly the same as) $H(z)$ measurements, leading to different sensitivities the model parameters. For this particular model, the strong degeneracies imply that small values of $\lambda$ can't be well constrained by cosmology data alone. One can also note that the corresponding posterior distribution of $\lambda$ for the current data is quite non-Gaussian, the simulated future data merely enhances this effect.

As expected the impact on the equation of state $w$ is more modest---about a factor of two---since in this class of models the CMB temperature is not directly sensitive to it. It is also interesting to note that we now obtain a tight constraint on the matter density even without using the Planck prior---indeed this constraint, which improves the current one by about a factor of seven, is comparable to (though slightly larger than) that prior.

\section{Forecasts for previously constrained models}
\label{previous}

Constraints on three classes of cosmological models for which the standard temperature-redshift relation is violated have recently been provided by \cite{Gelo}, using the same background cosmology and CMB data being used in our present work. In this section we discuss how these current constraints are improved by the simulated future data, again with the goal of quantifying the gains in sensitivity to be expected from the next generation of astrophysical facilities, and how these gains may depend on the parameter space of each model.

In each of the following three subsections we start by a concise introduction to each model, and then compare the current constraints obtained in \cite{Gelo} with our forecasts.
Our description of these models is by no means exhaustive: our goal is merely to highlight the feature of each of them (such as the Einstein equations) which and relevant for our analysis and/or for a comparison to other models, and we the reader to the original works for additional details on each model. We also note that in this section no Planck prior is used, and that we will again neglect the contribution of the radiation density to the Friedmann equation, since we are only considering low-redshift data.

\subsection{A decaying dark energy model}
\label{oldjetzer}

This model was developed by Jetzer {\it et al.} \cite{Jetzer1,Jetzer2} based on earlier work in \cite{Lima}. It is effectively a decaying dark energy model, where continuous photon creation is possible, which obviously implies a non-standard temperature-redshift relation. In flat FLRW universes and with the previously mentioned assumptions, the Friedmann equation is
\be
E^2(z)=\frac{3\Omega_m-m}{3-m}(1+z)^3+\frac{3(1-\Omega_m)}{3-m}(1+z)^m\,.
\ee
where $m$ is a dimensionless model parameter, which can equivalently be expressed as an effective dark energy equation of state
\be
w_{eff}=\frac{m}{3}-1\,.
\ee
The model includes a coupling to radiation, which must necessarily be weak \cite{Chluba}, as otherwise it would be already ruled out, so at a phenomenological level we may consider it a small perturbation on the standard $\Lambda$CDM behaviour. These assumptions lead to the temperature-redshift relation
\be
T(z)=T_0(1+z)^{3w_r}\left[\frac{(m-3\Omega_m)+m(\Omega_m-1)(1+z)^{m-3}}{(m-3)\Omega_m}\right]^{w_r}\,.
\ee
where we have also relaxed the assumption of a strict radiation fluid, by allowing for a radiation-type fluid with a generic (but still assumed constant) equation of state $w_r$. Clearly, with $m=0$ and $w_r=1/3$ we recover the $\Lambda$CDM case, so this model falls into the first of the model classes discussed in the introduction. Unlike the model in the previous section, here the temperature-redshift relation does depend on the matter density, but this dependency is quite mild---as we will see presently, those on $m$ and $w_r$ are much stronger.

\begin{figure*}
\begin{center}
\includegraphics[width=0.32\textwidth]{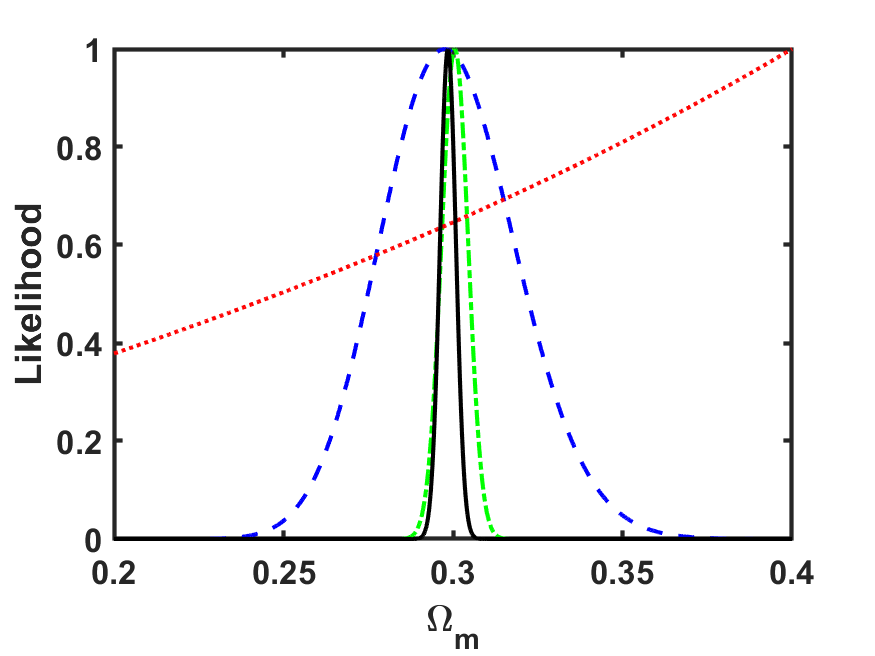}
\includegraphics[width=0.32\textwidth]{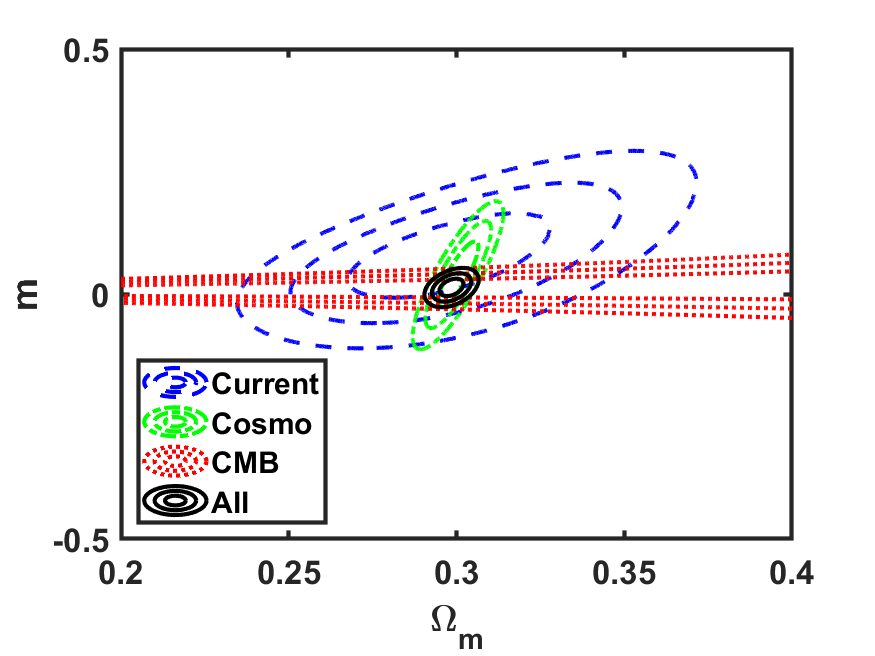}
\includegraphics[width=0.32\textwidth]{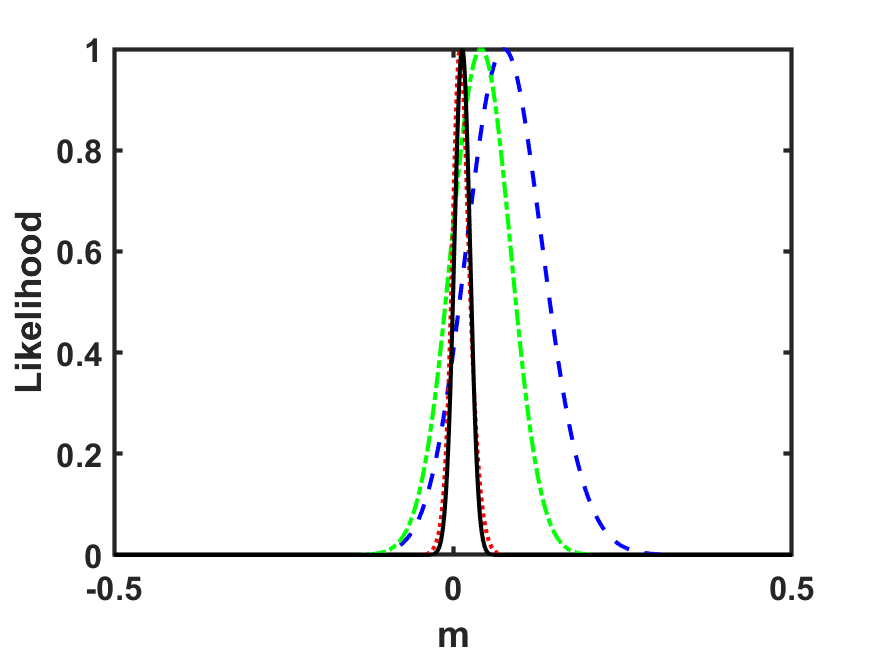}
\end{center}
\caption{Forecast constraints on the decaying dark energy model with $w_r=1/3$. The middle panel shows the one, two and three sigma constraints in the two-dimensional $\Omega_m$--$m$ parameter space, and the side panels show the one-dimensional (marginalized) posterior likelihoods for each of the parameters. Green dash-dotted, red dotted and black solid lines depict the cosmological, CMB and combined forecasts for our simulated future data. For comparison, the dashed blue lines show the current constraints, already obtained in \cite{Gelo}.}
\label{figure04}
\end{figure*}
\begin{figure*}
\begin{center}
\includegraphics[width=0.49\textwidth]{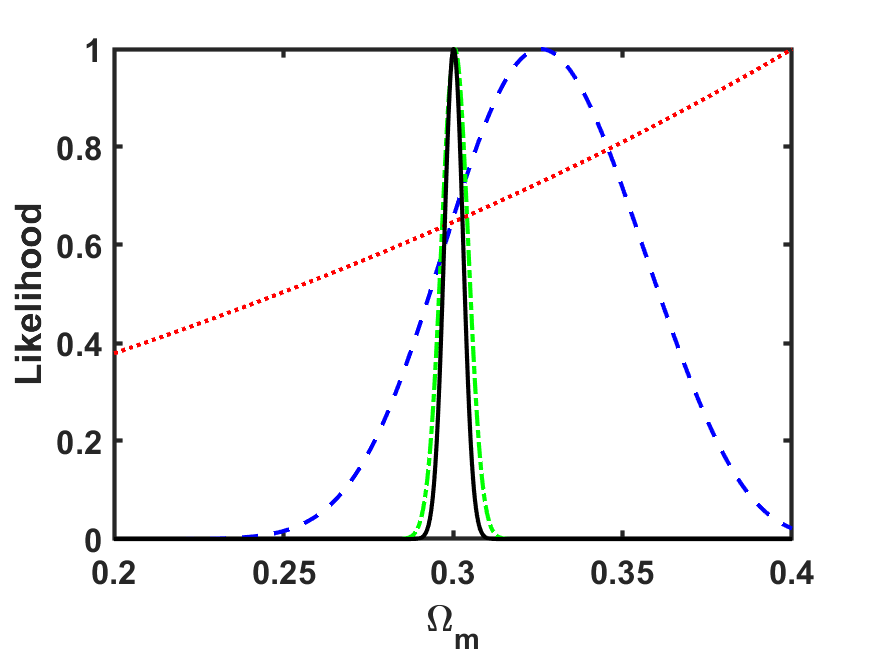}
\includegraphics[width=0.49\textwidth]{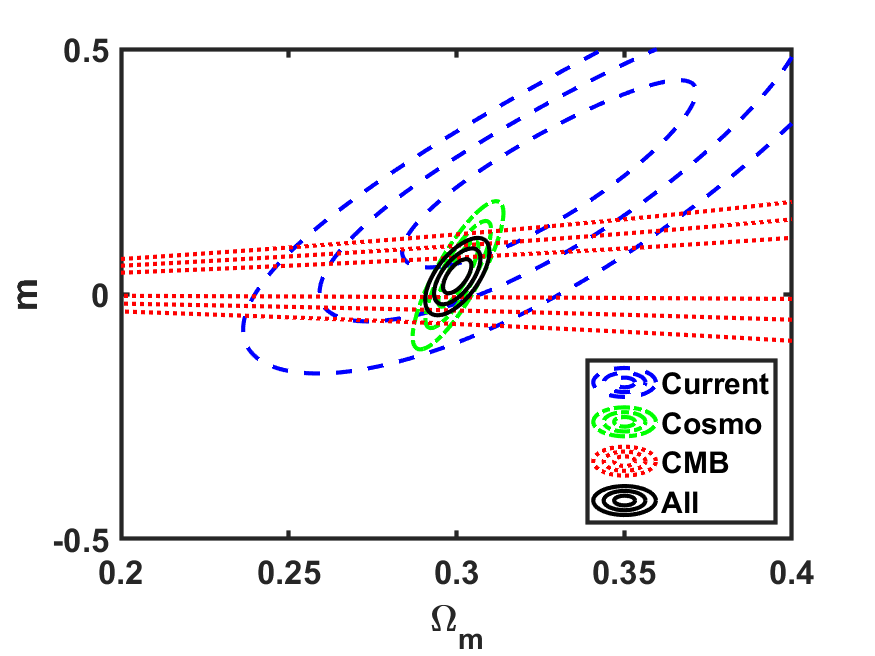}
\includegraphics[width=0.49\textwidth]{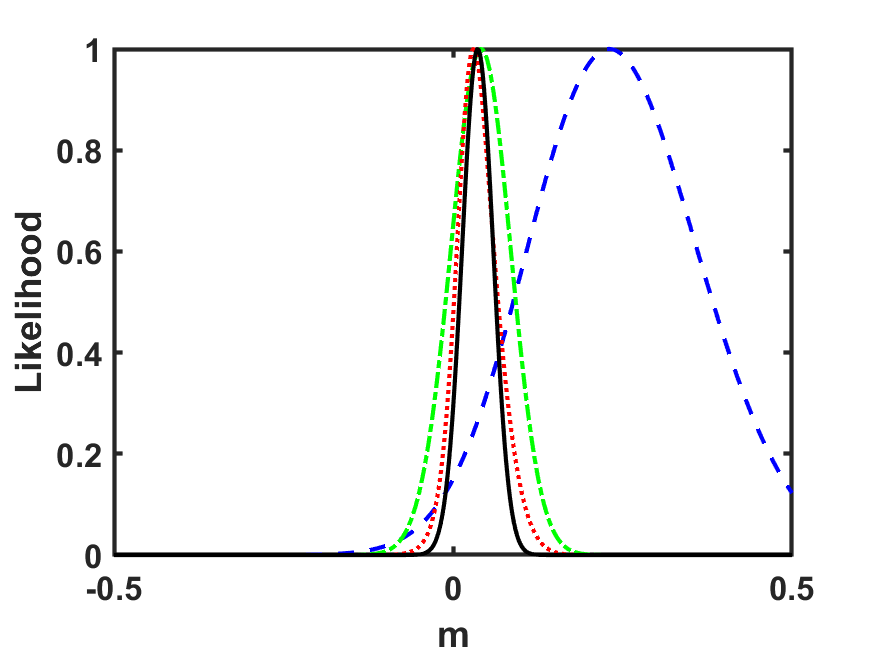}
\includegraphics[width=0.49\textwidth]{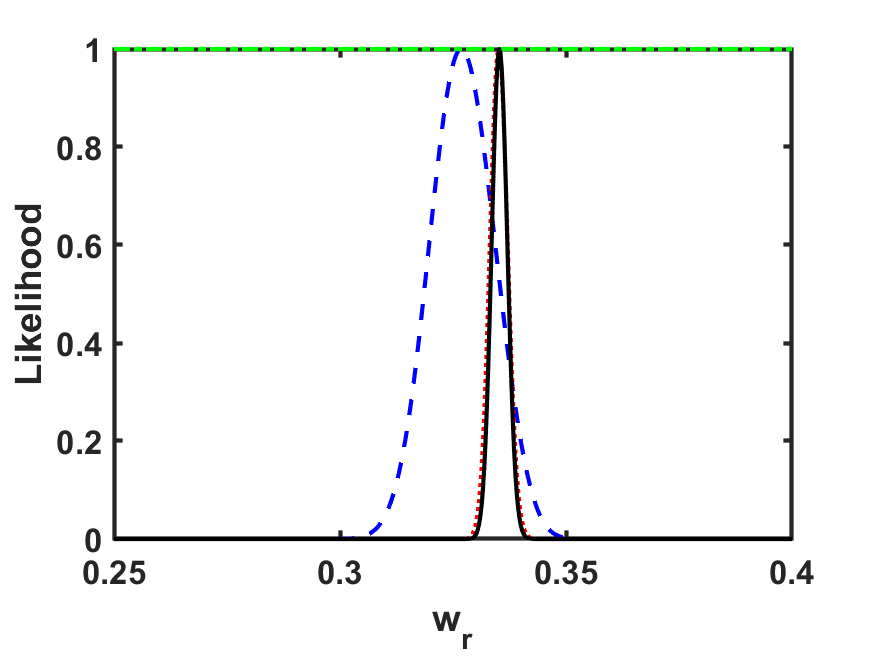}
\end{center}
\caption{Same as Figure \ref{figure04} for a more general equation of state of radiation $w_r$, assumed to be constant. The one-dimensional (marginalized) posterior likelihood for $w_r$ is also shown.}
\label{figure05}
\end{figure*}

The analysis of \cite{Gelo} reports, for the case with standard radiation ($w_r=1/3$). the one-sigma constraints
\bq
\Omega_m&=&0.30\pm0.02\\
m&=&0.07\pm0.05\,,
\eq
while if one relaxes the $w_r=1/3$ assumption and includes this as a third free parameter, the constraints become
\bq
\Omega_m&=&0.33\pm0.03\\
m&=&0.23\pm0.13\\
w_r&=&0.35\pm0.02\,.
\eq
Overall, there are no statistically significant deviations from the $\Lambda$CDM behaviour. A simpler analysis, assuming a fixed matter density, $\Omega_m=0.315$, can be found in \cite{Riechers}.

Figure \ref{figure04} summarizes our forecast results for the $w_r=1/3$ case, while Figure \ref{figure05} shows the results for the case where $w_r$ is an additional free parameter. In both figures, the blue dashed lines depict the current constraints, obtained by \cite{Gelo}. For the former case we find the one-sigma constraints
\bq
\Omega_m&=&0.299\pm0.002\\
m&=&0.013\pm0.011\,,
\eq
while for the latter the constraints become
\bq
\Omega_m&=&0.300\pm0.003\\
m&=&0.03\pm0.02\\
w_r&=&0.335\pm0.002\,.
\eq
This model's parameter space is slightly different from the one for the torsion model, but one common point is that the CMB temperature has very little sensitivity to the matter density but high sensitivity to the other parameters, thus breaking the degeneracy between each of them and $\Omega_m$. In this case, our simulated data would significantly improve the constraints on all three parameters, with respect to the current constraints: by a factor of ten for the matter density and $w_r$, and by a factor of about five for $m$. This would also translate into sub-percent level constraints on the effective equation of state dark energy, which are compatible with those expected from contemporary cosmological probes assuming more standard cosmological models.

\subsection{A scale invariance model}
\label{oldcanuto}

This model was developed by Canuto {\it et al.} \cite{Canuto1,Canuto2} and stems from the notion that although the effects of scale invariance must vanish in the presence of particles with non-zero rest masses, one may assume that on cosmological scales empty space could still be scale invariant. A mathematical formulation of this idea leads to a bimetric type theory, with a time-dependent function $\lambda$ playing the role of a scale transformation factor between the ordinary matter frame and a separate scale invariant frame. The first of these frames can be thought of as the atomic (or physical) frame, while the second is a gravitational frame, in which the ordinary Einstein equations would still hold \cite{Canuto2}.

For a flat FLRW universe, with the further simple but natural assumption of a generic power-law function of the form $\lambda(t)=(t/t_0)^p=x^p$, where $t_0$ is the present age of the universe (and we have defined a dimensionless age of the universe, $x$), the Friedmann equation is
\be
\left(E(z,x)+\frac{p}{2x}\Omega_\lambda\right)^2=\Omega_m(1+z)^{3(1+w)}x^{-p(1+3w)}+\Omega_{\Lambda}x^{2p}\,,
\label{scalef2}
\ee
where we have also introduced a dimensionless parameter $\Omega_\lambda=2/(t_0H_0)$. This shows that the present age of the universe is also a free parameter of the model. Note that there is a consistency condition $(1+p\Omega_\lambda/2)^2=\Omega_m+\Omega_{\Lambda}$, and that setting $\lambda(t_0)=1$ ensures that $\Lambda$CDM is recovered for $p=0$.

For numerical convenience this can be re-written as
\bq
E(z,x)&=&\frac{\Omega_\lambda}{2x}\left[-p+\sqrt{N(z,x)}\right]\\
\frac{\Omega_\lambda^2}{4}N(z,x)&=&\Omega_m(1+z)^{3(1+w)}x^{2-p(1+3w)}+\Omega_\Lambda x^{2(1+p)}\,,
\eq
together with
\be
\frac{dx}{dz}=- \frac{x}{1+z}\times\frac{1}{\sqrt{N(z,x)}-p}\,,
\ee
with the initial condition $x=1$ at $z=0$. Finally, the temperature-redshift relation is
\be
T(z,x)=T_0(1+z)x^{-p/2}\,,
\ee
which depends only on $p$, but not on the matter density. As in the torsion case, this is an example of a model in the second of the classes mentioned in the introduction. Setting $\Omega_\Lambda=0$ would make it a genuine alternative to $\Lambda$CDM (i.e., there would be no parametric limit which recovers it), but previous work \cite{Covariant} has shown that such a scenario is ruled out. Therefore, we limit ourselves to the case where $\Omega_\Lambda\neq0$, in which the model is a parametric extension of $\Lambda$CDM.

\begin{figure*}
\begin{center}
\includegraphics[width=0.32\textwidth]{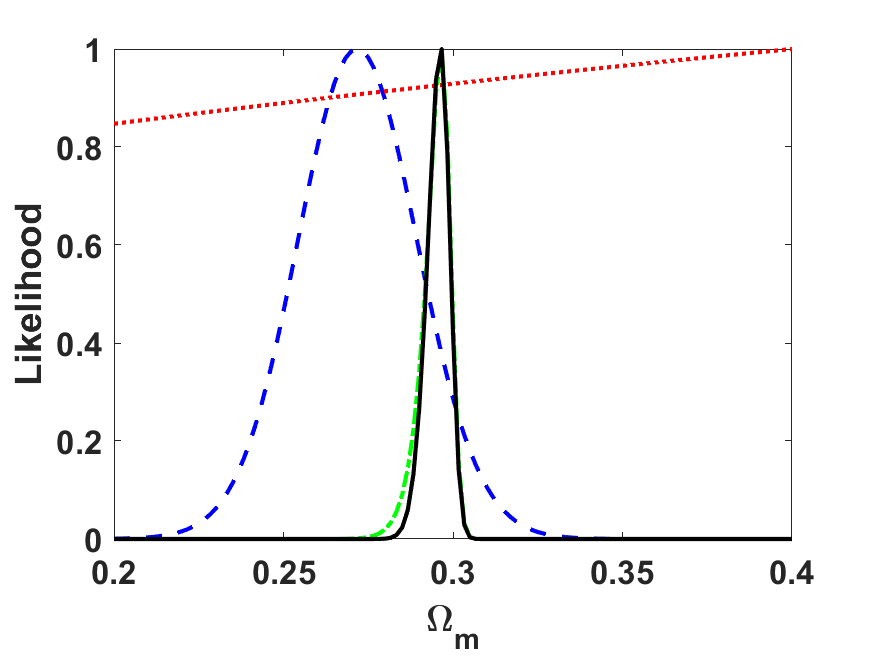}
\includegraphics[width=0.32\textwidth]{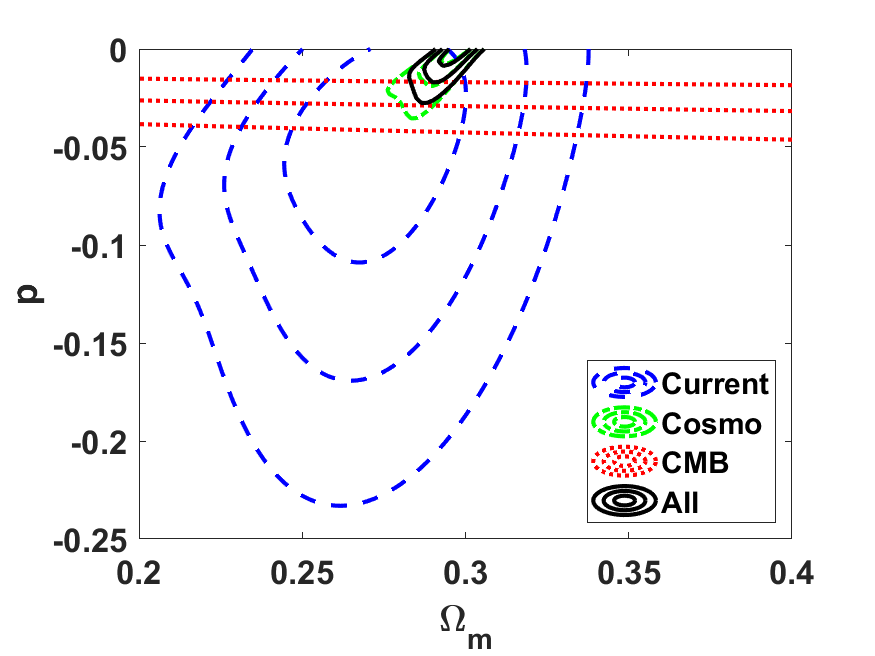}
\includegraphics[width=0.32\textwidth]{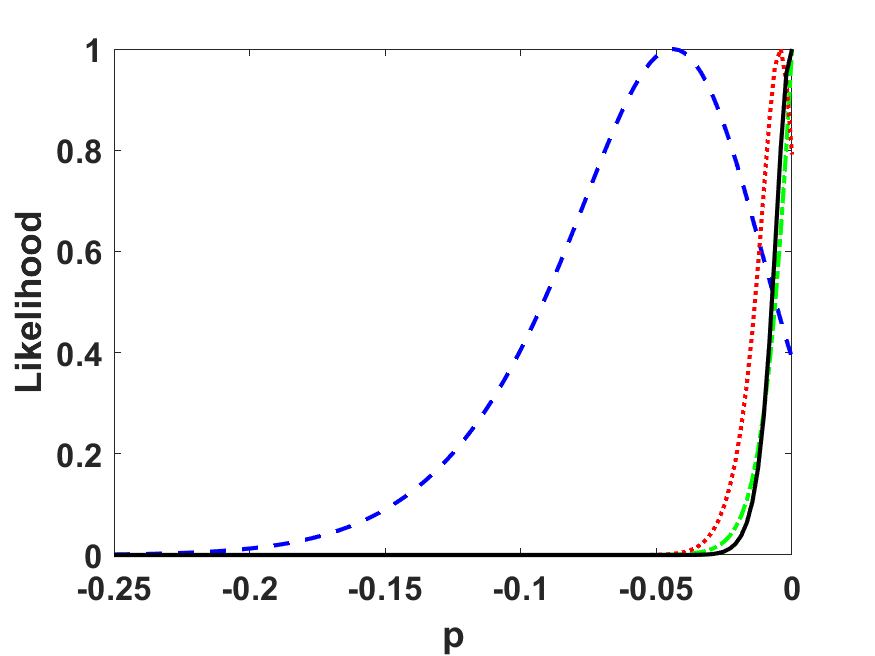}
\end{center}
\caption{Forecast constraints on the scale invariance model. The middle panel shows the one, two and three sigma constraints in the two-dimensional $\Omega_m$--$p$ parameter space, and the side panels show the one-dimensional (marginalized) posterior likelihoods for each of the parameters. Green dash-dotted, red dotted and black solid lines depict the cosmological, CMB and combined forecasts for our simulated future data. For comparison, the dashed blue lines show the current constraints, already obtained in \cite{Gelo}.}
\label{figure06}
\end{figure*}

This model was constrained using the same current data as we are using \cite{Gelo}, under the assumption of a standard equation of state for matter\footnote{Including $w$ as an additional parameter is not warranted in a statistical sense, since without it the model already overfits the data.}, $w=0$, and a uniform prior on $\Omega_\lambda$ corresponding to a present age of the universe from 13.5 to 27 Gyr. The choice of the lower limit corresponds to the age of the oldest identified galaxy, GN-z11 \cite{Eleven}, further corroborated by estimates from galaxy clusters \cite{Valcin}. The analysis was also restricted to the physically more plausible case $p\le0$, in which case the model effectively has a decaying cosmological constant, and therefore shares some qualitative similarities with the model in the previous subsection. With these assumptions \cite{Gelo} obtained
\bq
\Omega_m&=&0.27\pm0.02\\
p&>&-0.14\,,
\eq
the first being a one-sigma constraint and the second a two-sigma lower limit. As in the previous model, in this case the CMB temperature data does not significantly constrain the matter density, but it does help to break degeneracies between the model's parameters.

To make the comparison with the constraints obtained in the earlier work reliable, we retain all the above assumptions, which in any case do not significantly impact the results. Figure \ref{figure06} shows the results of our forecast analysis using the simulated data, which leads to the improved constraints
\bq
\Omega_m&=&0.297\pm0.003\\
p&>&-0.0135\,.
\eq
This improves the matter density constraint by about a factor of six, and the lower limit on the power-law dependence of the scale transformation factor $p$ by about a factor of ten.

\subsection{The fractional cosmology model}
\label{oldfractional}

The third model previously studied in \cite{Gelo} is the fractional cosmology model \cite{Fractional}, which relies on the mathematical formalism of fractional calculus \cite{Tarasov}. Conceptually, the later can be thought of as a parametric extension of the usual concepts of differentiation and integration. As in the case of the model in the previous sub-section, this was was originally envisaged as an alternative to $\Lambda$CDM, in the sense that it is assumed to contain no cosmological constant. However, \cite{Gelo} have shown that this scenario is ruled out, precisely due to the CMB temperature data: the best-fit values of the model parameters obtained separately from cosmology and from CMB temperature date are mutually incompatible at many standard deviations. Therefore, in what follows we only address the model's most favourable scenario, which includes a cosmological constant and also fixes the present age of the universe to its standard value\footnote{As in the previous sub-section, including $t_0$ as an additional parameter is not warranted in a statistical sense, since without it the model already overfits the data.}.

For a flat FLRW universe, the Friedmann equation for this model has the form
\be
E^2+\frac{1-\alpha}{H_0t_0} \frac{E}{x}=\Omega_m(1+z)^{3}x^{\alpha-1}+\Omega_\Lambda\,,
\label{fraceq1}
\ee
where $t_0$ and $x$ have the same definitions as in the previous sub-section, and $\alpha$ is the fractional calculus parameter: in standard calculus one has $\alpha=1$. Importantly, since $E(z=0)=1$, we have 
\be
\frac{1-\alpha}{H_0t_0}=\Omega_m+\Omega_\Lambda-1\,;
\ee
thus the requirement of a positive age of the universe implies that for $\Omega_m+\Omega_\Lambda<1$ one needs $\alpha>1$; conversely, $\alpha<1$ would require $\Omega_m+\Omega_\Lambda>1$. 

The mathematical similarities between this model and the one in the previous subsection should be clear, and they also extend to the fact that there is a numerically more convenient expression for the Friedmann equation. Fixing the age of the universe to the standard value, we can write
\be
E(z,x)=\frac{\alpha-1}{2x}+ \sqrt{\frac{(\alpha-1)^2}{4x^2}+\Omega_m(1+z)^3x^{\alpha-1}+2-\alpha-\Omega_m}\,,
\ee
together with
\be
\frac{dx}{dz}=-\,\frac{1}{(1+z)E(z,x)}\,.
\ee
Finally, in this model the temperature redshift relation is
\begin{equation}
    T(z,x)=T_0(1+z)x^{(\alpha-1)/3}\,,
\end{equation}
all the above reduce to the standard flat $\Lambda$CDM behaviour for $\alpha=1$. Once again, note that the temperature-redshift relation is only sensitive to the beyond-$\Lambda$CDM parameter $\alpha$.

The earlier analysis, using the data in Sect. \ref{nowdata}, led to the following one-sigma constraints
\bq
\Omega_m&=&0.28\pm0.02\\
\alpha&=&1.05\pm0.04\,.
\eq

\begin{figure*}
\begin{center}
\includegraphics[width=0.32\textwidth]{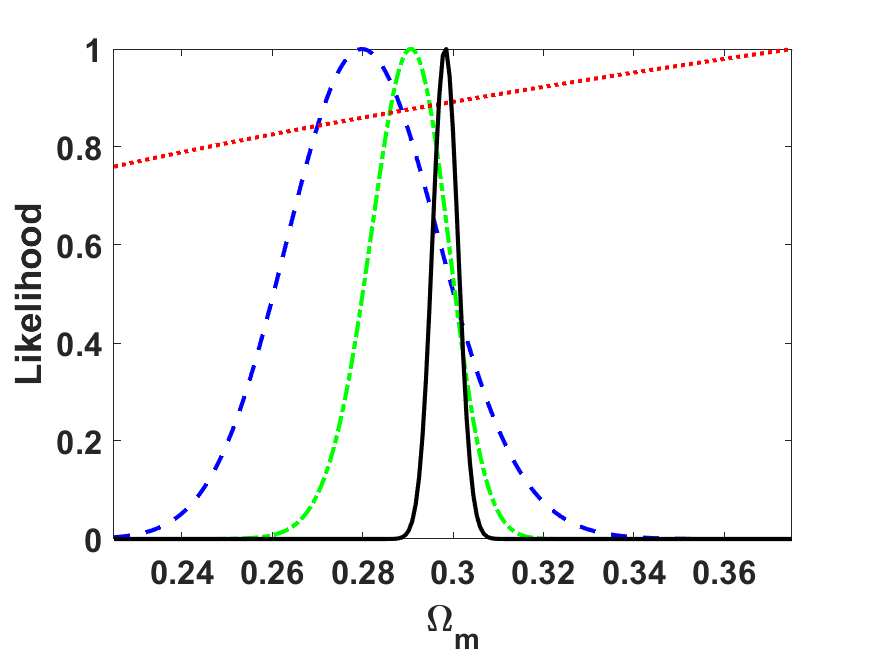}
\includegraphics[width=0.32\textwidth]{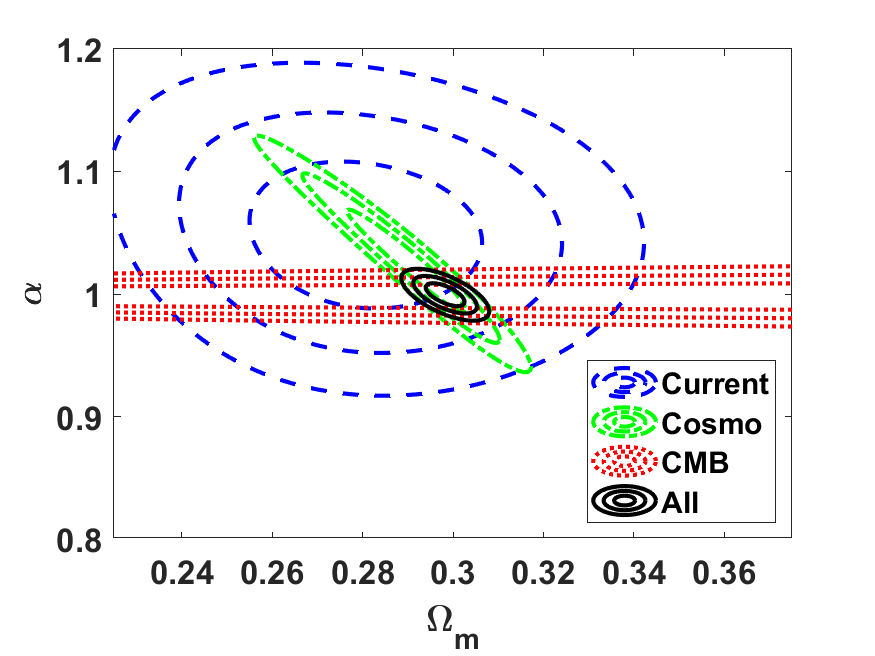}
\includegraphics[width=0.32\textwidth]{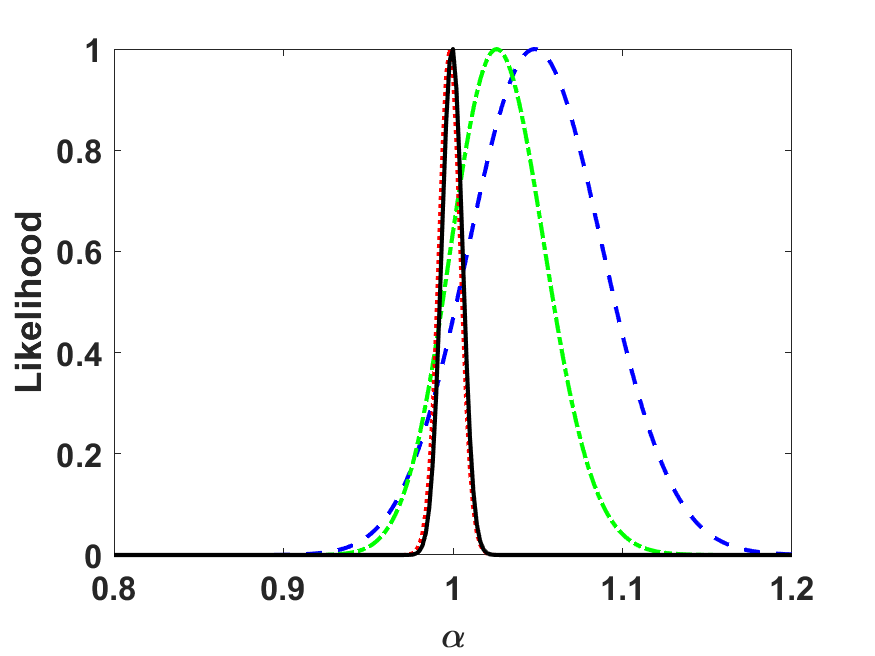}
\end{center}
\caption{Forecast constraints on the fractional cosmology model. The middle panel shows the one, two and three sigma constraints in the two-dimensional $\Omega_m$--$\alpha$ parameter space, and the side panels show the one-dimensional (marginalized) posterior likelihoods for each of the parameters. Green dash-dotted, red dotted and black solid lines depict the cosmological, CMB and combined forecasts for our simulated future data. For comparison, the dashed blue lines show the current constraints, already obtained in \cite{Gelo}.}
\label{figure07}
\end{figure*}

Figure \ref{figure07} shows the results of our forecast for this model, for which we obtain the constraints
\bq
\Omega_m&=&0.299\pm0.003\\
\alpha&=&1.00\pm0.02\,.
\eq
Qualitatively the behaviour is similar to that of the previous models, though the gains in sensitivity for each of the model parameters are quantitatively different. More specifically, in this case the constraint on the matter density is increased by a factor of six, while that on $\alpha$ is only improved by a factor of two.

\section{A further dynamical dark energy model}
\label{newmodel}

We finally consider a dynamical dark energy model recently proposed in \cite{Gupta}. This is a phenomenological model, in the sense that it cannot be derived from an action, but it turns out to also have some similarities with the models discussed in the previous section. The model stems from the possibility, discussed in \cite{Gomide}, of a time varying temporal metric coefficient described by a function $f(t)$ which can be thought of as analogous to the scale factor $a(t)$, and therefore it could also be thought of as a particular type of varying speed of light model \cite{VSL}. It is meant to be an alternative to the standard cosmological model, in the sense that a cosmological constant is not expected. Considering as usual flat FLRW models, and retaining the assumption of \cite{Gupta} on the analytic form of the function,
\be
f(t)=e^{\alpha(t-t_0)}\,,
\ee
one obtains the following Friedmann equation
\be
\left(H+\alpha\right)^2=\frac{8\pi G}{3}\rho\,;
\ee
note that here the model parameter $\alpha$ (not to be confused with the fine-structure constant) has dimensions of inverse time. Further assuming fluids with constant equations of state $w=p/\rho=const$, the continuity equation leads to
\be\label{scaling}
\rho\propto a^{-3(1+w)} e^{-(1+3w)\alpha(t-t_0)}\,.
\ee
For a low-redshift universe dominated by ordinary matter ($w=0$), and further defining the two dimensionless parameters $\beta=\alpha t_0$ and $\epsilon=H_0t_0$, we can write the Friedmann equation in the numerically convenient form
\be
E(z,x)=1+\sqrt{\Omega_m}\left[(1+z)^{3/2}e^{-\frac{\beta}{2}(x-1)}-1\right]\,,
\ee
together with
\be\label{timez}
\frac{dx}{dz}=-\,\frac{1}{(1+z)\epsilon E(z,x)}\,.
\ee
Additionally, the temperature-redshift relation has the form
\begin{equation}\label{ttz}
T(z,x)=T_0(1+z)e^{-\frac{\beta}{2}(x-1)}\,.
\end{equation}
Note that in this model allows for a non-standard age of the universe, determined by the parameter $\epsilon$. Structural phenomenological similarities of this model with the scale invariance and fractional cosmology models discussed in the previous section should be clear: specifically, visual inspection of the Einstein equation in all three cases immediately shows their similar forms. In passing we also note that in this model the standard distance-duality relation is also violated.

\begin{figure*}
\begin{center}
\includegraphics[width=0.32\textwidth]{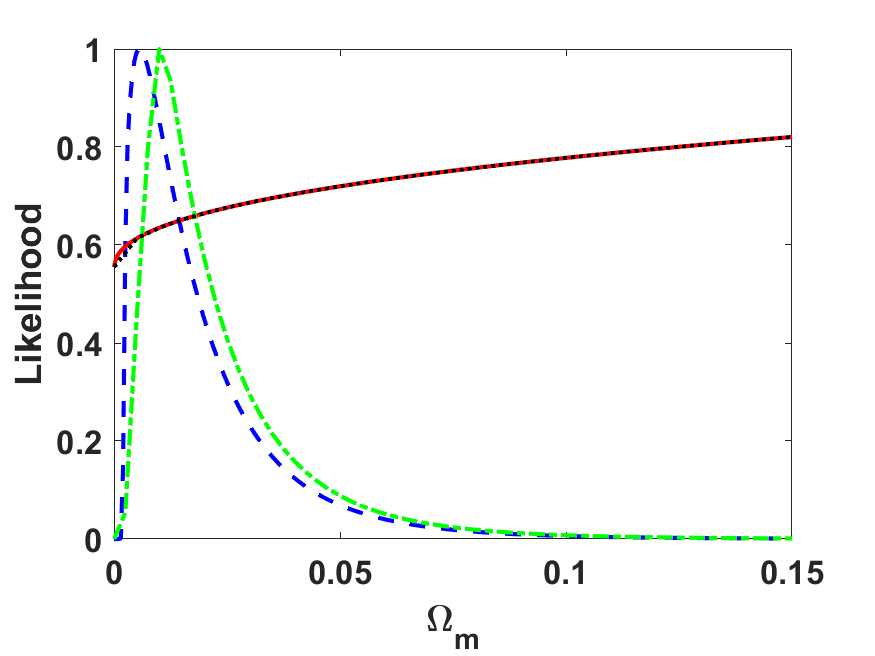}
\includegraphics[width=0.32\textwidth]{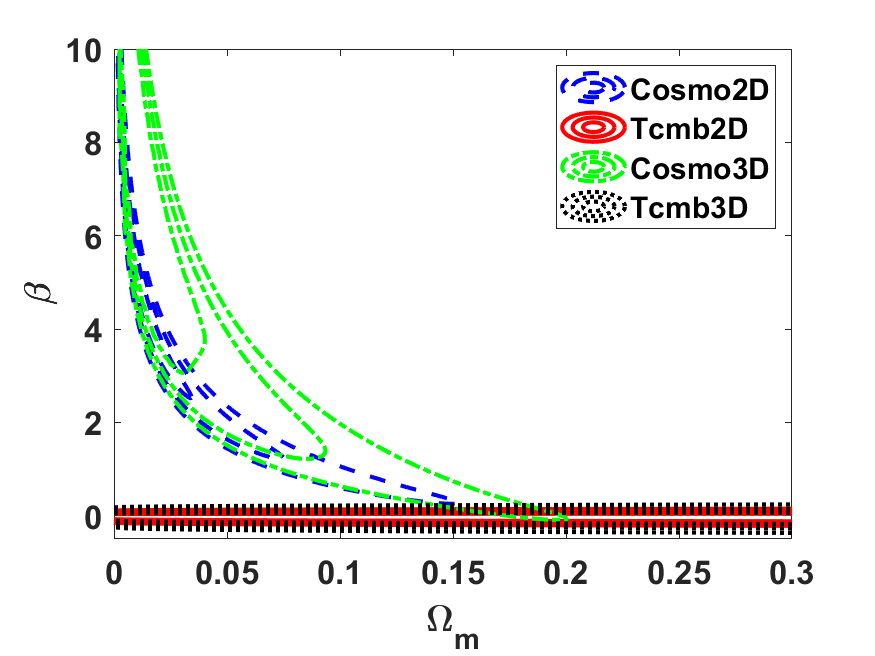}
\includegraphics[width=0.32\textwidth]{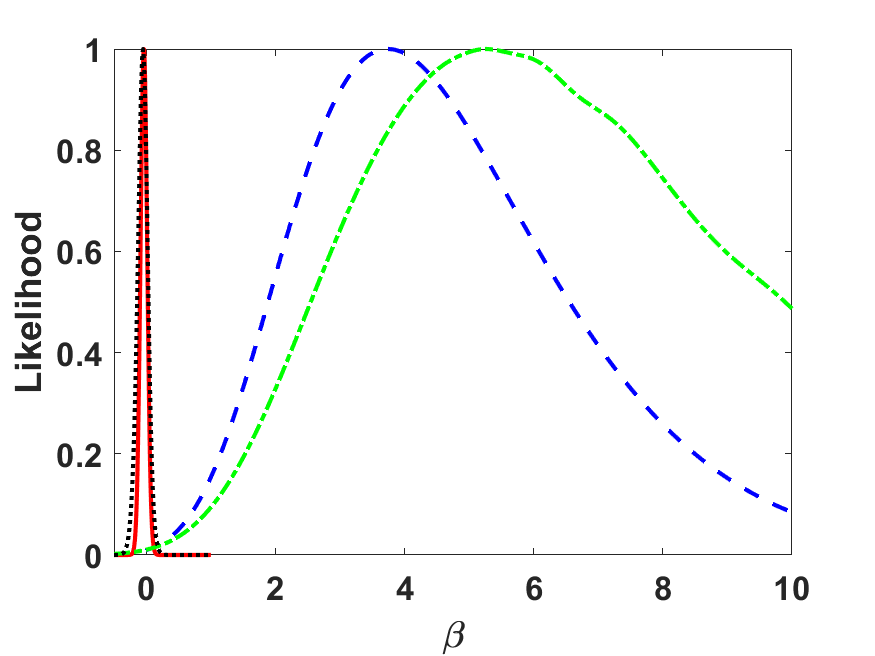}
\end{center}
\caption{Current constraints on the Gupta model, for $\Omega_\Lambda=0$. The middle panel shows the one, two and three sigma constraints in the two-dimensional $\Omega_m$--$\beta$ parameter space, and the side panels show the one-dimensional (marginalized) posterior likelihoods for each of the parameters. The constraints are shown separately for the cosmological and CMB data, and 2D and 3D denote the cases with the age of the universe having the standard value and being an additional free parameter, respectively.}
\label{figure08}
\end{figure*}

\subsection{Without a cosmological constant}

We start with the baseline case without a cosmological constant, and also assume a standard age of the universe, in which case we have a two-dimensional parameter space. Figure \ref{figure08} shows the results of our analysis. One notices that a degeneracy between the two parameters exists for the cosmological data, while the CMB data is only sensitive to $\beta$. Importantly, the constraints from the cosmological and CMB data, which are shown separately, are mutually incompatible, implying that the model is ruled out (and that a combination of the two data sets is not meaningful). Specifically, the cosmology data yields the constraints
\bq
\Omega_m&=&0.006_{-0.003}^{+0.010}\\
\beta&=&3.74_{-1.68}^{+2.31}\,.
\eq
The preferred matter density is of course problematic on its own, being much smaller than the baryon density, but the constraint on $\beta$ also conflicts with the one obtained from the CMB data, which is
\be
\beta=-0.04_{-0.06}^{+0.05}\,.
\ee

One may also extend the model, allowing for a non-standard age of the universe. This introduces a third free parameter in the model, which can then be marginalized. As was done for the fractional cosmology model in the previous section, for this age we assume a uniform prior from 13.5 to 27 Gyr.  Figure \ref{figure08} also depicts the results of this analysis, which does not lead to any major changes. The error bars on the model parameters are slightly increased (as expected) but the best-fir values are not significantly changed, and the model is still ruled out. In this case the cosmology data leads to
\bq
\Omega_m&=&0.010_{-0.004}^{+0.009}\\
\beta&=&5.28_{-2.37}^{+3.65}\,.
\eq
while the CMB temperature gives
\be
\beta=-0.03_{-0.10}^{+0.05}\,.
\ee
As for the age of the universe, there is a very mild (and not statistically significant) preference for larger values, but the parameter remains unconstrained.

\begin{figure*}
\begin{center}
\includegraphics[width=0.32\textwidth]{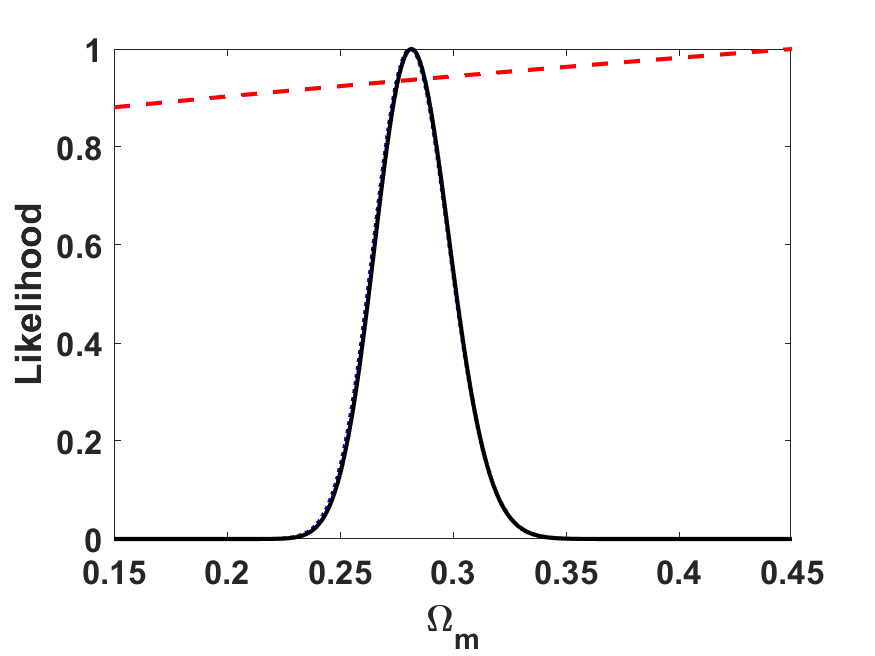}
\includegraphics[width=0.32\textwidth]{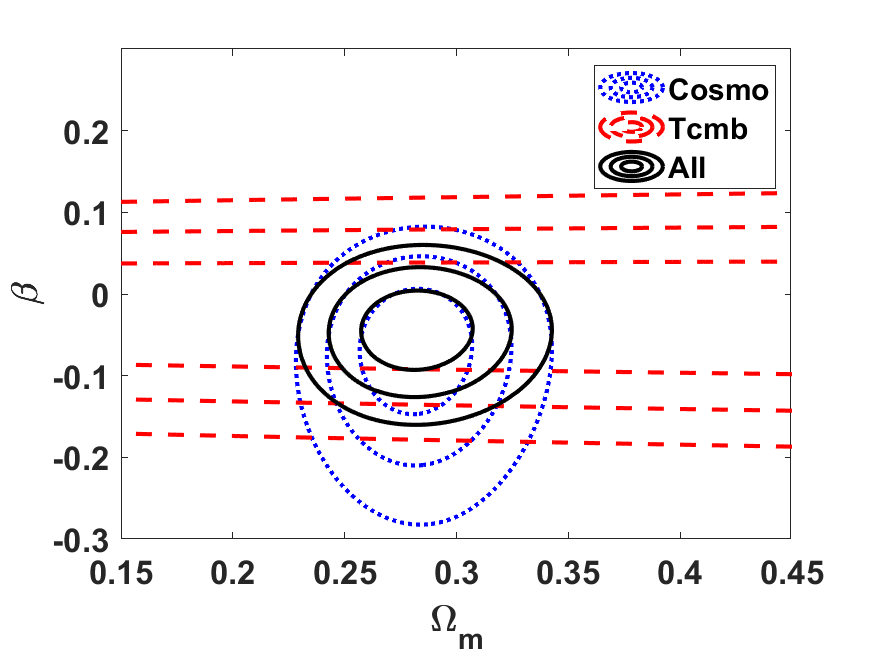}
\includegraphics[width=0.32\textwidth]{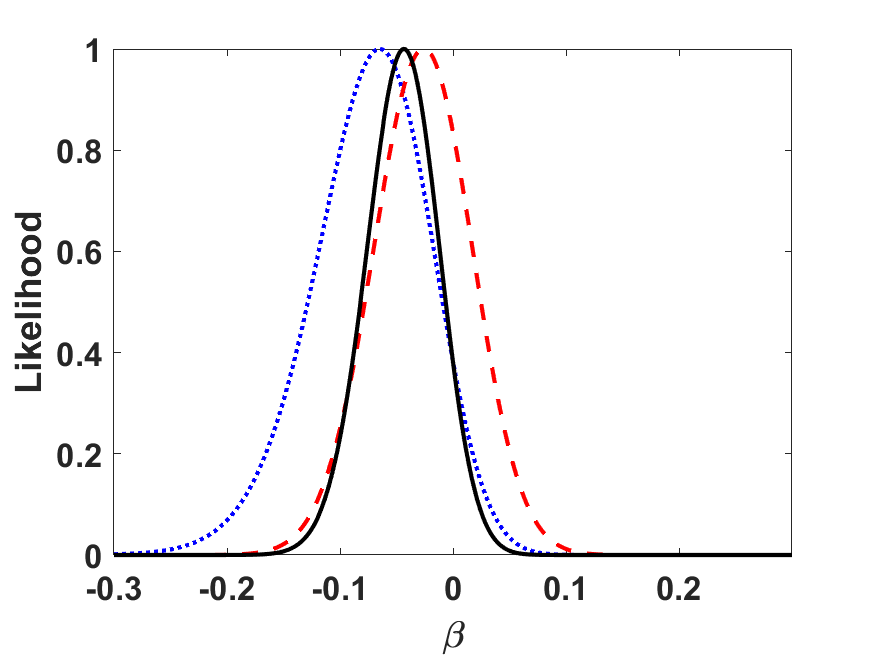}
\end{center}
\caption{Current constraints on the Gupta model, including a cosmological constant and with a standard value of the age of the universe. The middle panel shows the one, two and three sigma constraints in the two-dimensional $\Omega_m$--$\beta$ parameter space, and the side panels show the one-dimensional (marginalized) posterior likelihoods for each of the parameters. Blue dotted, red dashed and black solid lines depict the cosmological, CMB and combined constraints respectively.}
\label{figure09}
\end{figure*}

\subsection{Including a cosmological constant}

Given that the model cannot be a genuine alternative to the standard one, we may ask how it behaves if one allows for the presence of a cosmological constant, which again would make the model a parametric extension of $\Lambda$CDM. Clearly this is a purely phenomenological addition, but it is nevertheless interesting to see how much such a model is allowed (by current or future data) to deviate from $\Lambda$CDM.

According to Eq.(\ref{scaling}), for a fluid with an equation of state $w=-1$, its density will evolve in this model as
\be
\rho_\Lambda\propto e^{2\alpha(t-t_0)}\,.
\ee
Including this term, and further assuming flatness, the Friedmann equation can be rewritten,
\be
\left[E(z,x)+\frac{\beta}{\epsilon}\right]^2=\Omega_m(1+z)^3e^{-\beta(x-1)}+\left[\left(1+\frac{\beta}{\epsilon}\right)^2-\Omega_m\right]e^{2\beta(x-1)}\,,
\ee
while Eqs.(\ref{timez}--\ref{ttz}) are still valid. It is clear that flat $\Lambda$CDM is recovered for $\beta=0$ and $\epsilon=1$.

Figure \ref{figure09} shows our resulting constraints, assuming a standard age for the universe. The two data sets are now mutually consistent, and as expected the results are fully consistent with $\Lambda$CDM. The one-sigma combined constraints are
\bq
\Omega_m&=&0.28\pm0.02\\
\beta&=&-0.04\pm0.03\,.
\eq
It is also worthy of note that in this case the cosmology and CMB data have, on their own, comparable constraining power on $\beta$, though as usual only the former constrains the matter density.

We have also verified that allowing the age of the universe to be a further independent parameter does not have a significant impact, other than slightly relaxing the constraints on the parameter $\beta$. Specifically, in this case we find
\bq
\Omega_m&=&0.28\pm0.02\\
\beta&=&-0.05\pm0.06\,.
\eq
As in the case without a cosmological constant, there is a modest and not statistically significant preference for larger than standard values of the universe's age.

\begin{figure*}
\begin{center}
\includegraphics[width=0.32\textwidth]{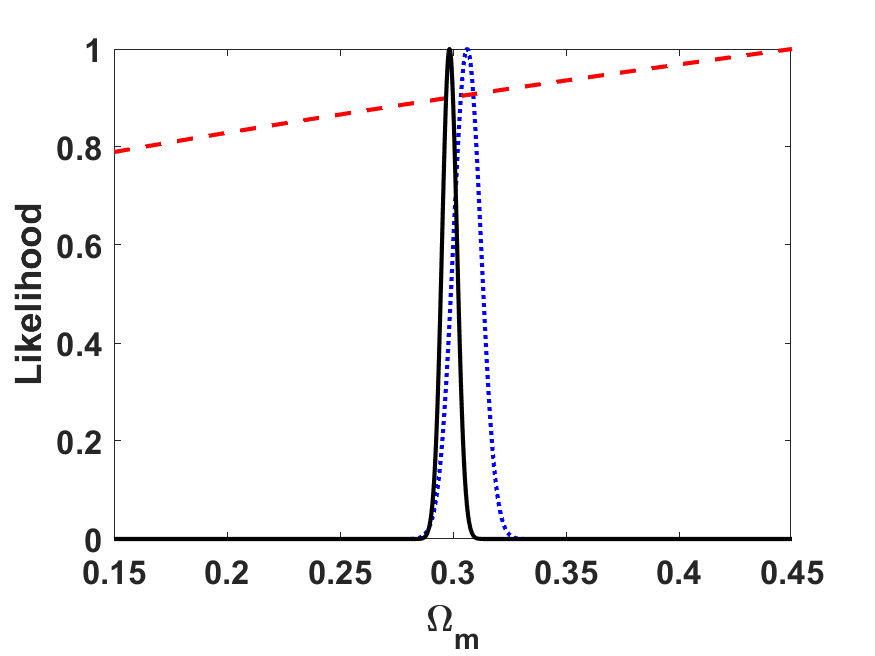}
\includegraphics[width=0.32\textwidth]{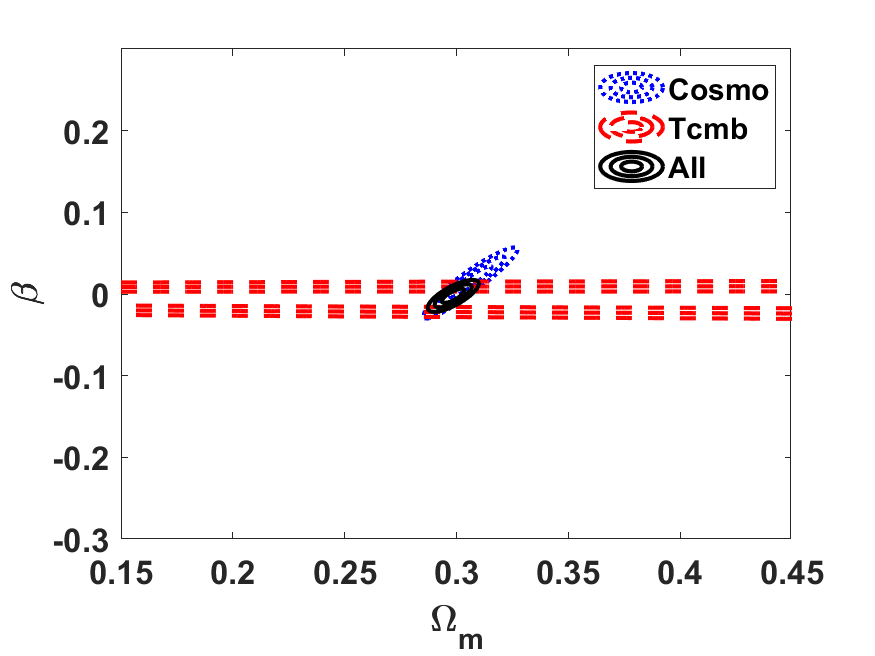}
\includegraphics[width=0.32\textwidth]{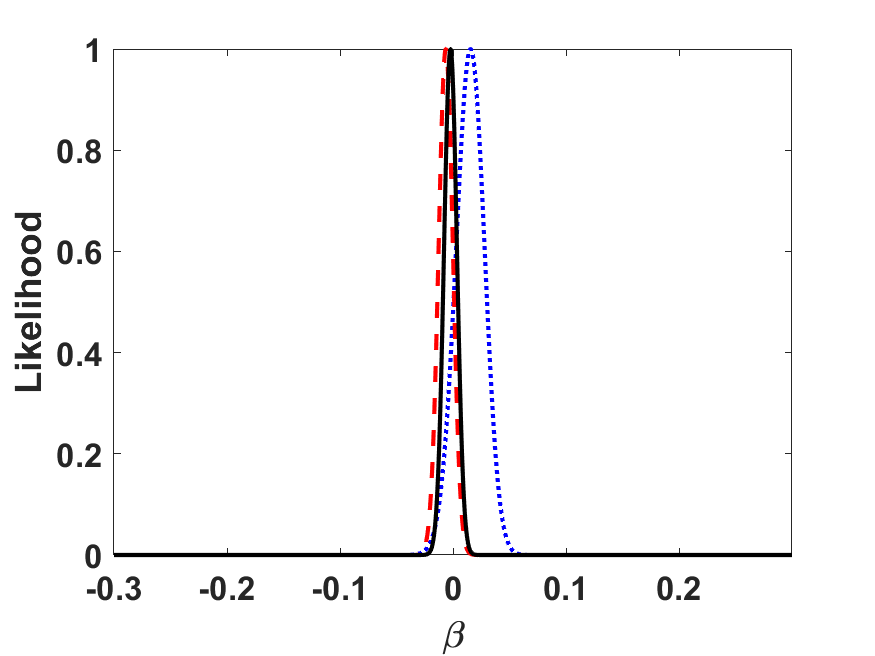}
\end{center}
\caption{Same as Figure \ref{figure09}, for the simulated future data discussed in the text.}
\label{figure10}
\end{figure*}

Finally, we discuss the constraints on this model expected from the future simulated data. We restrict ourselves to the case with a cosmological constant and the standard age of the universe, for the reasons already explained. These results are presented in Fig. \ref{figure10}, to be compared with the current constraints in Fig. \ref{figure09}.

As expected we again find significant gains in sensitivity. The one-sigma combined constraints are
\bq
\Omega_m&=&0.298\pm0.003\\
\beta&=&-0.002\pm0.006\,.
\eq
corresponding to improvements by factors of about six and five respectively.

\section{Conclusions}
\label{concl}

We have extended the recent analysis by \cite{Gelo} of the cosmological impact of measurements of the CMB temperature at non-zero redshift, both by studying two additional models, steady-state torsion \cite{Torsion1,Torsion2} and a specific type of varying speed of light model \cite{Gupta}. Both of these stem from different physical assumptions than the models considered in \cite{Gelo}, although the second turns out, in practice, to be fairly similar to earlier models. We also provided forecasts of the gains in constraining power to be expected from next-generation facilities. These gains, which are summarized in Table \ref{table4}, are clearly model-dependent, but are typically of factors of a few to ten, for each of the model parameters. Still, it is worthy of note that there is always a significant improvement in the constraints on the matter density.

We note that in our current constraints analysis we have chosen to use the same background cosmology datasets as the previous works we build upon. In principle these could be slightly updated, but that would make the comparison to the previous works less easy (and also somewhat less fair), while improvements might only be modest. For example, replacing Pantheon by Pantheon+, the improvement in the final constraints would be of order ten percent or less.

\begin{table}
\begin{center}
\caption{Summary of the expected gains, for the models considered in this work, in the sensitivity of the one sigma posterior constraint on the matter density and on the beyond-$\Lambda$CDM model parameter(s), for the simulated future data with respect to the current constraints.}
\label{table4}
\begin{tabular}{c c c}
\hline
Model & $\Omega_m$ Gain  & Other parameter gains \\
\hline
Torsion (no prior) & 7 & 6 $(\lambda)$, 2 $(w)$ \\
Jetzer {\it et al.} & 10 & 5 $(m)$, 10 $(w_r)$ \\ 
Canuto {\it et al.} & 6 & 10 $(p)$ \\ 
Fractional & 6 & 2 $(\alpha)$ \\ 
Gupta (with $\Lambda$) & 6 & 5 $(\beta)$ \\ 
\hline
\end{tabular}
\end{center}
\end{table}

Our analysis confirms the earlier results, showing that there is a wide class of models for which this data has a constraining power that is comparable to that of other low-redshift background cosmology data. At the broad conceptual level, the main reason for this is that constraints from cosmological data exhibit significant degeneracies between model parameters---typically including the matter density and one or more parameters quantifying the model's deviation from the standard $\Lambda$CDM behaviour. In this wide class of models, the temperature-redshift relation is modified and depends predominantly (or, in some cases, even exclusively) on the new model parameters. The CMB data therefore leads to stringent constraints on such parameters, thereby breaking the aforementioned degeneracies.

Admittedly, our forecasts rely on a number of simplifying assumptions regarding next-generation ground and space based astrophysical facilities which at present are in various different stages of development. The SKA and Roman Space Telescope are under construction, with first light foreseen for ca. 2027, while the ANDES spectrograph is currently in Phase B and CORE is simply a mission concept at this stage. Forecasts for the redshift drift are particularly uncertain at this point since the signal has not been detected so far, though it should be detectable by both ANDES and the SKA. The first dedicated redshift drift experiment is presently ongoing, using the ESPRESSO spectrograph, and its first results (expected in the coming months) will provide an important assessment of the feasibility of future detections. In any case, our results show that temperature measurements will remain a competitive probe of new physics, and motivate efforts for their continuing improvement.

\section*{Acknowledgements}

This work was financed by Portuguese funds through FCT (Funda\c c\~ao para a Ci\^encia e a Tecnologia) in the framework of the project 2022.04048.PTDC (Phi in the Sky, DOI 10.54499/2022.04048.PTDC). CJM also acknowledges FCT and POCH/FSE (EC) support through Investigador FCT Contract 2021.01214.CEECIND/CP1658/CT0001 (DOI 10.54499/2021.01214.CEECIND/CP1658/CT0001).

\bibliographystyle{model1-num-names}
\bibliography{tcmb}
\end{document}